%% file: ms.tex
\documentclass[sigconf,nonacm]{acmart}

\usepackage{graphicx}
\usepackage{subfiles}
\usepackage{soul,color}

\usepackage{todonotes}
\usepackage{booktabs}
\PassOptionsToPackage{hyphens}{url}\usepackage{hyperref}

\usepackage{dsfont}
\usepackage{amsmath}
\usepackage{nicefrac}

\usepackage{amsthm}

\theoremstyle{definition}
\newtheorem{definition}{Question}

\usepackage{CJKutf8}

\usepackage{caption}
\usepackage{subcaption}
\usepackage{tablefootnote}
\usepackage{pifont}
\newcommand{\cmark}{\textcolor{teal}{\ding{51}}}%
\newcommand{\xmark}{\textcolor{red}{\ding{55}}}%
\newcommand\ml[1]{\textcolor{teal}{#1}}

\usepackage{multicol}
\usepackage{multirow}
\usepackage{url}

\usepackage{tikz}
\tikzset{
  treenode/.style = {shape=rectangle, rounded corners,
                     draw, align=center,
                     top color=white, bottom color=white},
  root/.style     = {treenode, font=\Large, bottom color=red!30},
  env/.style      = {treenode, font=\normalsize},
  dummy/.style    = {circle,draw}
}

\usepackage{xcolor}

\newcommand{\clirtask}{CLIR}

\AtBeginDocument{%
  \providecommand\BibTeX{{%
    \normalfont B\kern-0.5em{\scshape i\kern-0.25em b}\kern-0.8em\TeX}}}

\begin{document}

\title[Overview of the TREC 2023 NeuCLIR Track]{Overview of the TREC 2023 NeuCLIR Track}
\author{Dawn Lawrie,$^\dagger$ Sean MacAvaney,$^\ddagger$ James Mayfield,$^\dagger$ \\ 
Paul McNamee,$^\dagger$ Douglas W. Oard,${^o}^\dagger$ Luca Soldaini,$^*$ Eugene Yang$^\dagger$}
\affiliation{
  \institution{$^\dagger$Johns Hopkins University Human Language Technology Center of Excellence,\\
  $^\ddagger$University of Glasgow, $^o$University of Maryland, $^*$Allen Institute for AI}
  \country{}
}

\email{lawrie@jhu.edu,sean.macavaney@glasgow.ac.uk,mayfield@jhu.edu}
\email{mcnamee@jhu.edu, lucas@allenai.org, oard@umd.edu, eugene.yang@jhu.edu}

\renewcommand{\shortauthors}{Lawrie et al.}

\begin{abstract}

The principal
goal of the TREC Neural Cross-Language Information Retrieval (NeuCLIR) track
is to study the impact of neural approaches to cross-language information retrieval.
The track has created four collections,
large collections of Chinese, Persian, and Russian newswire
and a smaller collection of Chinese scientific abstracts.
The principal tasks are ranked retrieval of news in one of the three languages, using English topics.
Results for a multilingual task, also with English topics but with documents from all three newswire collections,
are also reported.
New in this second year of the track is a pilot technical documents CLIR task
for ranked retrieval of Chinese technical documents using English topics.
A total of 220 runs across all tasks were submitted by six participating teams
and, as baselines, by track coordinators.
Task descriptions and results are presented.
\end{abstract}

\settopmatter{printfolios=true}
\maketitle

\input{1-intro} %

\section{News Retrieval Tasks}

In this section we describe the \clirtask\ and MLIR tasks.

\input{clir-2-task} %
\input{clir-3-docs} %
\input{clir-4-topic} %
\input{clir-5-judgment} %
\input{clir-6-resource} %
\input{clir-7-participation} %
\input{clir-8-results} %

\section{Technical Documents Pilot Task}

\input{tech-2-task} %
\input{tech-3-docs} %
\input{tech-4-topic} %
\input{tech-5-judgment} %
\input{tech-6-resource} %
\input{tech-7-participation} %
\input{tech-8-results} %

\input{9-future} %
\input{10-conclusion} %

\section{Acknowledgments}
We would like to thank Victor Lavrenko, Cynthia Wu, and Mahsa Yarmohammadi for their help with translations.

\bibliographystyle{ACM-Reference-Format}
\bibliography{biblio}

\appendix

\input{_figs_overlap}

\input{_table_clir_full_results}

\input{_table_mlir_full_results}
\input{_table_tech_full_results}

\end{document}

%% file: 1-intro.tex
\section{Introduction}

This is the second year of the TREC Neural Cross-Language Information Retrieval (NeuCLIR) track.\footnote{The NeuCLIR track website is at \url{https://neuclir.github.io}}
The first year of the track included CLIR tasks for news, with twelve participating teams~\cite{lawrie2022overview}.
This second year of the track continues those tasks, and adds two new tasks:
(1) Multilingual Information Retrieval (MLIR);
and (2) a new technical document CLIR pilot task.
This overview describes the track's five tasks and reports results.

The TREC NeuCLIR track was developed in response to the confluence of three factors.
First, neural algorithms have demonstrated ranked retrieval effectiveness that is substantially improved over prior techniques,
but the application of neural techniques in CLIR has not yet been fully characterized.
Second, computational infrastructure such as tokenization and embedding are now available for many languages,
but CLIR poses additional challenges that have not yet been fully addressed.
Third, earlier TREC-style CLIR test collections are small in comparison to current monolingual ranked retrieval test collections,
and when pooling was used (as was almost always the case) the relevance judgments for those collections had been developed by pooling only pre-neural CLIR models.
Each of the NeuCLIR track's tasks seeks to address aspects of this challenge.

The track's principle tasks are ranked CLIR for news,
with topics in English\footnote{The news test collections also provide topics in Chinese, Persian and Russian, so queries in languages other than English can also be constructed.}
and news documents in one of three languages (Chinese, Persian, or Russian).
These are the most mature of the track's tasks, and the capabilities that they provide are foundational to the other two tasks.
Task participants face two broad classes of challenge
that distinguish CLIR from monolingual retrieval:
(1) less robust training data than is presently available for monolingual ranked retrieval tasks;
and (2) imbalances and misalignments in present multilingual embeddings that must be addressed
to optimize the use of that computational infrastructure for CLIR tasks.
Monolingual ranked retrieval results,
created using topics in the collection language, are also reported as a baselines.
Five of the six participating teams submitted CLIR runs.

New in this second year of the track is ranked MLIR for news,
with topics in English.
This task requires generating a single ranked list for a given topic
that includes Chinese, Persian and Russian documents.
 The principal additional challenge in this task is that scores computed for documents
 in different languages are usually incomparable,
 making generation of a unified ranked list difficult.
 While data from NeuCLIR 2022 can be used to evaluate MLIR algorithms,
 there were no runs submitted for MLIR in 2022.
 Three of the six participating teams submitted results for the MLIR task in 2023.

Also new in this second year of the track is a CLIR pilot task on technical documents.
The goal of this task is to search a collection of Chinese dissertation abstracts using English topics.
We characterized this as a pilot task because the technical nature of the topics
and the documents calls for specialized assessor expertise,
so in this pilot task we are as interested in the details of test collection construction
as we are in the results obtained by participating teams.
The additional challenge for participating teams
is to accommodate technical vocabulary in a CLIR setting,
for which existing computational infrastructure may be less well suited than it is for news.
Five of the six participating teams submitted results for the Technical Documents CLIR pilot task.

The news test collections in Chinese, Persian and Russian are the same as in the TREC 2022 NeuCLIR track,
but new topics were developed for this second year of the track.
Building on lessons learned from the first year of the track,
these new topics were designed to optimize their utility for evaluation in the MLIR task. 
Of the 76 topics created during topic development,
62 Chinese topics, 60 Persian topics, 62 Russian topics, and 65 MLIR topics were retained for evaluation of systems.
Relevance judgments for 40 Technical Documents CLIR topics were used for evaluation runs. %

The remainder of this paper is organized as follows.
We begin with a brief summary of the CLIR and MLIR tasks,
emphasizing changes since TREC 2022;
readers can refer to the TREC 2022 NeuCLIR track overview paper in the published proceedings
for additional details~\cite{lawrie2022overview}. %
This is followed the results of those tasks.
Next, we present full details and complete results for the new Technical Documents CLIR pilot task.
The NeuCLIR track will continue in 2024,
so we follow our presentation of task design and results with a section on our initial thoughts about future directions for the track.  Finally, we conclude with some remarks on what we have learned from this year's track.

%% file: clir-2-task.tex
\subsection{Task Definitions}

We have two news retrieval tasks, \clirtask\ and MLIR.
The \clirtask\ task includes three ``settings'' (i.e., sub-tasks):
ad hoc CLIR, reranking CLIR, and monolingual retrieval.
All three settings 
use the same document collections, topics, and relevance assessments.
Monolingual runs use topics manually translated into the language of the documents;
ad hoc and reranking runs use the original English topics.
Ad hoc runs rank documents from the entire collection,
while reranking runs rank only the 1,000 documents that appear in the output of a NIST-provided initial run.

\subsubsection{Ad Hoc CLIR Setting}
\label{subsub:clir-ad-hoc}
The first setting 
in the NeuCLIR track is ad hoc CLIR.
Systems receive a document collection in Chinese, Persian, or Russian,
and a set of topics written in English.
For each topic, the system must return a ranked list of 1,000 documents
drawn from the entire target language document collection,
ordered by likelihood and degree of relevance to the topic.
Runs that use a human in the loop for ad hoc retrieval
(or had design decisions influenced by human review of the topics)
are indicated as ``manual'' runs;
all others are considered ``automatic.''

\subsubsection{Reranking CLIR Setting}
The reranking setting 
provides an initial ranked list of 1,000 retrieved documents from the document collection.
Each ranked list is the output of a BM25 retrieval system,
which used document translation to cross the language barrier.
The run IDs are 
{\texttt{*-hltcoe-MTES-patapscoBM25RM3td}}, where
 * is \texttt{zho} for Chinese, \texttt{fas} for Persian, and \texttt{rus} for Russian. 
The runs appear in bold in Tables~\ref{tab:zho-full-results}, \ref{tab:fas-full-results}, and \ref{tab:rus-full-results}.
These runs use the concatenation of the English title and description for each topic as the query for that topic. 
Systems are then asked to rerank the documents to produce a new ordering that improves an
evaluation metric.
This setting is suitable for teams that want to focus on second-stage scoring models,
rather than on models that search an entire collection.

\subsubsection{Monolingual Setting}%
While monolingual retrieval is not a focus of the NeuCLIR track,
monolingual runs can improve assessment pools
and serve as points of reference for cross-language runs.
The monolingual retrieval setting is identical to the ad-hoc setting,
but it uses topic files that are human translations of the English topics
into a target language in a way that would be expressed by native speakers of the language.
This setting is suitable for teams looking to explore monolingual ranking in languages other than English.
It also has a lower barrier to entry than the cross-language tasks.

\subsubsection{Multilingual Information Retrieval (MLIR) Task}

In NeuCLIR 2023, a new multilingual retrieval task was added. 
This task is identical to the \clirtask\ task described in \S\ref{subsub:clir-ad-hoc},
except systems must search all three document collections and produce a single unified ranked list. 
In other words, systems should treat the three document collections (\S\ref{sub:documents})
across all three languages as a single corpus. 
Participants were informed that, since topics for this task are derived from those of the \clirtask\ task,
there is no guarantee that each topic will have relevant results in each language.

%% file: clir-3-docs.tex
\subsection{Documents}
\label{sub:documents}

NeuCLIR 2023 continues to use the NeuCLIR~1 document set,
which was also used for NeuCLIR 2022.
The collection consists of roughly 2 million Persian documents, 3 million Chinese documents, and almost 5 million Russian documents
spanning the years 2016 to 2021.
For more information about how to extract the text from Common Crawl News documents
and how the collection can be obtained, see the NeuCLIR 2022 Overview paper~\cite{lawrie2022overview}. 

%% file: clir-4-topic.tex
\begin{table*}[]
\caption{Relevance judgment statistics for NIST topics and Track Coordinator topics. }\label{tab:judgment-nist-coordinators}
    \centering
    
\begin{tabular}{l|cc|cc|cc|cc}
\toprule
& \multicolumn{2}{c|}{Chinese} & \multicolumn{2}{c|}{Persian} & \multicolumn{2}{c|}{Russian} & \multicolumn{2}{c}{MLIR} \\
          Topic Developer &    NIST & Coordinators &    NIST & Coordinators &    NIST & Coordinators &    NIST & Coordinators \\
\midrule
\# Topics Retained          &      38  &      24 &      37 &       23 &      38 &      24  &  38 & 27\\
Avg. \# Judgments / Topic &  357.61 &  353.08 &  345.00 &  334.13 &  327.29 &  317.71 &  1,020.82 & 880.81 \\
Avg. Prevalence           &   6.83\% &   4.05\% &   7.98\% &   3.27\% &   9.86\% &   8.37\% & 8.18\% & 5.18\%\\
\bottomrule
\end{tabular}

\end{table*}

\subsection{Topics}
\label{sec:topics}

NeuCLIR 2022 topics were developed as traditional TREC-style information needs
that are broader than cross-language question answering.\footnote{In cross-language question answering, the typical goal is to find a single short answer that, for example, consists of a phrase or a sentence.}
This year new approaches to topic development were explored
with the goal of creating a greater percentage of topics that contained relevant documents in multiple languages
than was achieved in 2022.  

Topics 200-247 were developed by NIST assessors and used one development process,
while topics 248-275 were developed by the Track Coordinators 
using a different process.
Both cases prioritized creation of topics for which relevant documents in at least two of the three languages were relevant.

The process used for topic development by NIST assessors
paired two assessors with language skills in two different languages.
In a virtual meeting, the assessors brainstormed a topic together.
Good topics were described as topics that ``revolve around events, people, and places,
and be %
significant enough to have coverage in more than one language.''
After exploring the collection with monolingual searches in the two assessor languages,
a description was written,
followed by a first draft of the narrative,
and finally the title.
Then a single %
monolingual search was initiated in each language,
and each assessor counted the number of relevant documents in the top 25.
They were instructed to revise the topic if the count was less than one or greater than 20 in either of the languages.
Once the topic was appropriately scoped, the narrative was revised and the topic was included in the topics for 2023.

Because of unforeseen challenges,
NIST assessors were not available to create a sufficient number of topics.
Therefore, the Coordinators took over topic creation.
Since most of the Coordinators do not have language skills in any of the languages included in the collection,
a modified process for topic development was adopted.
A Coordinator worked independently to craft topics
following the general guidelines that the NIST assessors used with collection exploration,
leading to a title, description, and narrative.
Coordinators had access to two CLIR search engines:
a lexical retrieval system that indexed machine translated documents;
and a dense retrieval system implemented with ColBERT-X~\cite{colbert-x}.
The document display included the original document and a machine translation
(the same one provided to track participants; see Section~\ref{sub:documents}).
The browser could also be used to invoke Google Translate on the original document
should the user desire to do so. %
Coordinators could also use HiCAL~\cite{hical} to recommend documents to be judged.
Coordinators made relevance assessments using the same categories used during relevance judgment:
``very valuable,'' ``somewhat valuable,'' ``not that valuable,'' and ``not relevant.''
Coordinators were asked to judge at least 30 documents
and find between one and 20 documents in the very or somewhat valuable categories.
Document judgments were recorded.
Unlike the NIST-generated topics,
documents marked by coordinators as somewhat or very valuable %
were later added to the relevance judgment pools
that NIST assessors judged when making final relevance assessments. 

%% file: clir-5-judgment.tex
\subsection{Relevance Judgments}\label{sec:clir:rel-judgment}

Once all submissions were made,
by-language pools were created that integrated the top-ranked documents from both \clirtask\ and MLIR runs.
The top 50 documents for runs that teams prioritized as their top three runs.
Such runs have a checkmark in the JFD columns of Tables~\ref{tab:zho-full-results}, \ref{tab:fas-full-results}, and \ref{tab:rus-full-results}.
For other runs, a depth of 20 was used.
The Coordinators created some runs based on last year's system submissions as baselines that were also judged to a depth of 50.
The same 4-point scale as NeuCLIR22~\cite{lawrie2022overview} was used to judge relevance.
The 4-point scale was converted to a 3-point scale for the qrels,\footnote{The mapping from four relevance grades was 3->3, 2->1, 1->0, and 0->0. }
again as was done for NeuCLIR22~\cite{lawrie2022overview}.

After pooling, some topics were dropped from the \clirtask\ and MLIR tasks according to the following rules: 
\begin{itemize}
\item
If more than 40\% of the judged documents were judged to be somewhat or very valuable %
in a given language,
drop the topic from the \clirtask\ task in that language and from the MLIR task.
\item
If the relevance judgments for a topic have no documents in the somewhat or very valuable categories in a language,
drop that topic from the \clirtask\ task in that language,
but include it in the MLIR task if two languages have documents with at least one document in the somewhat or very valuable category per language. %
\end{itemize}
Of the 76 topics created, 62 were retained for each of Chinese and Russian, 60 for Persian, and 65 for the MLIR task.
All of the MLIR topics that were dropped were dropped because some language had too many relevant documents.
No topic had only relevant documents in a single language.
Table~\ref{tab:judgment-nist-coordinators} describes features of the topics that were retained
based on the criteria described in Section~\ref{sec:topics}.

Topics created by NIST assessors were removed for different reasons than those created by the Track Coordinators. 
Twelve of the fourteen problematic NIST assessor topics over all languages
(some topics had problems in multiple languages)
had too many relevant documents,
leading to their removal from the \clirtask\ and MLIR tasks.
By contrast, only one of the eleven problematic Coordinator topics had too many relevant documents.
For 75\% of the cases in which a NIST assessor topic had too many relevant documents in a language,
one of the two NIST assessors had checked for relevant documents in that language
using techniques that had been intended to minimize the occurrence of that problem. %
The two topics with no relevant documents in some language had not been checked by the NIST assessors in that language.
Of the languages for which assessors provided document counts for a topic during topic development, %
in only one-third of the cases were fifteen or more relevant documents discovered during topic development,
perhaps indicating that the BM25-based monolingual search system may have been a weaker retrieval system
than the CLIR systems submitted as runs.

Just over half of problematic %
Track Coordinator topics had three or fewer relevant documents.
Given that in all these cases NIST assessors judged the documents that the Coordinators had identified
as not that valuable or not relevant,
there are at least two possible reasons.
One is that the machine translations viewed by the Coordinators made a document appear relevant when it was not.
A second possibility is that the NIST assessor and the Coordinator disagreed on the definition of relevance.
The latter seems likely in cases where the Track Coordinators identified several relevant documents. 
This highlights the value of the usual preference for having the same assessor create the topic and judge the pools.

%% file: clir-6-resource.tex
\begin{figure*}[hbtp]
    \centering
    \includegraphics[width=\linewidth]{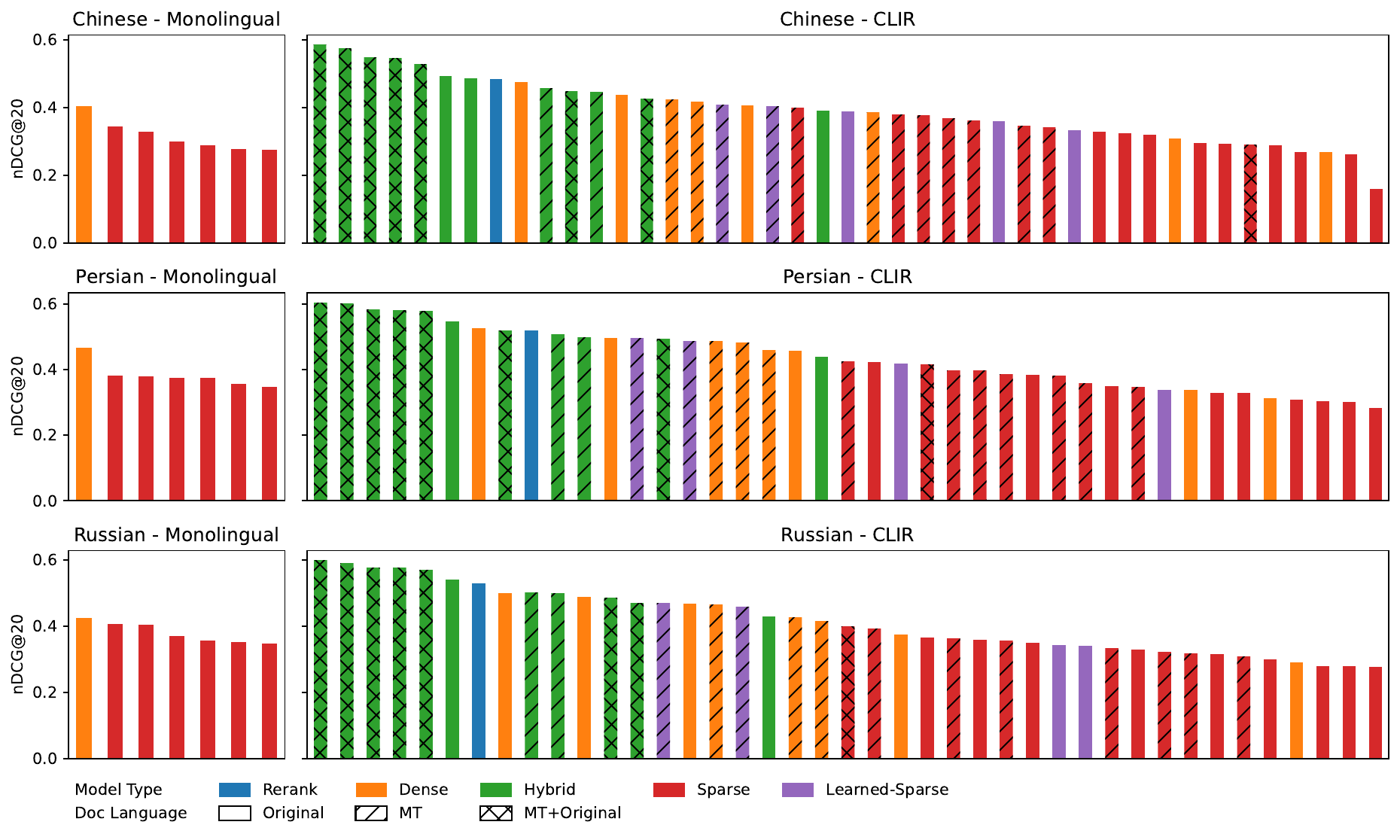}
    \caption{CLIR nDCG@20.}\label{fig:clir-ndcg-bar}
\end{figure*}

\subsection{Additional Resources}
\label{sec:clir-resources}
To support the aims of the \clirtask\ and MLIR tasks
and to lower the barrier to entry for new participants,
the track made available additional resources beyond the document collection and topics.
These resources included machine translated versions of queries and document collections,
translations of the widely used MS MARCO collection into the three NeuCLIR 2023 document languages,
and previously used IR test sets in the three NeuCLIR languages.

Machine translated versions of the queries were created by the online \textit{Google Translate} service.

English versions of the document collections were created by directional machine translation Transformer models
using the Sockeye version~2 toolkit.
As the document collection for the \clirtask\ and MLIR tasks did not change from 2022,
these are the same translations that were provided in the first year
and that are described in the TREC 2022 NeuCLIR Overview paper\cite{lawrie2022overview}.

As many neural IR systems are trained using data derived from the MS MARCO dataset\cite{bajaj2018ms},
translations of this English resource into different languages were provided.
We provided a version on Hugging Face called \textit{NeuMARCO}.\footnote{\url{https://huggingface.co/datasets/neuclir/neumarco}}
We also provided links to similar translations from the \textit{mMARCO} project\cite{bonifacio2022mmarco} on the NeuCLIR website.

The track website also collected a number of multilingual and bilingual resources in the languages of the track
including HC4 –- a CLIR collection built over three years of Common Crawl data in the same three languages~\cite{hc4},
as well as two multilingual CLIR datasets based on Wikipedia,
known as CLIRMatrix~\cite{clirmatrix} and WikiCLIR~\cite{wikiclir}.

Finally, the NeuCLIR 2022 topics and relevance judgments were available to track participants,
either from NIST, or in \texttt{ir\_datasets}.\footnote{\url{https://ir-datasets.com}}

\begin{figure*}[htp]
    \centering
    \includegraphics[width=\linewidth]{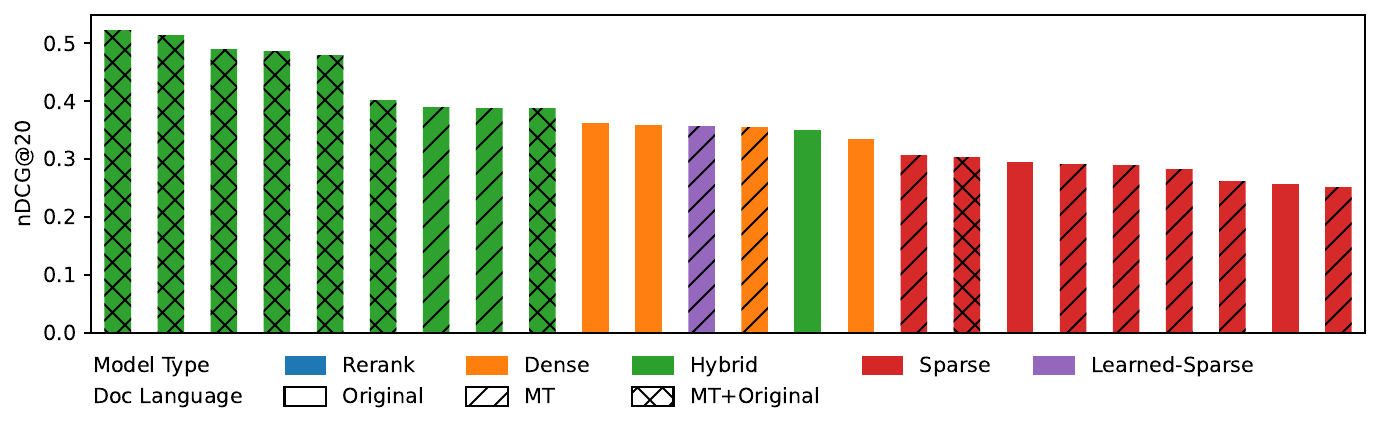}
    \caption{MLIR nDCG@20.}\label{fig:mlir-ndcg-bar}
\end{figure*}

%% file: clir-7-participation.tex
\subsection{Participation}

The main \clirtask\ task had five participants with 48 Russian runs, 48 Persian runs, and 49 Chinese runs:
\begin{itemize}
    \item Johns Hopkins University HLTCOE
    \item University of Maryland CLIP\footnote{NIST results show CLIP runs as HCIL, but the participating group at the University of Maryland was actually CLIP.}
    \item University of Massachusetts CIIR
    \item University of Southern California ISI
    \item University of Waterloo and Naver Labs
\end{itemize}
Three of these organizations also submitted to the MLIR task with 24 runs:
\begin{itemize}
    \item Johns Hopkins University HLTCOE
    \item University of Maryland CLIP
    \item University of Waterloo and Naver Labs
\end{itemize}

%% file: clir-8-results.tex
\subsection{Analysis of Official Submissions}

\begin{table*}[t]
\caption{Submitted Runs Summary.  Orig: Original language.}\label{tab:run-summary}
    \centering
\begin{tabular}{l|ccc|ccc|ccc|ccc}
\toprule
               & \multicolumn{3}{c|}{Chinese} & \multicolumn{3}{c|}{Persian} & \multicolumn{3}{c|}{Russian} & \multicolumn{3}{c}{MLIR} \\
Model Type \textbackslash~Doc Lang.      
               & Orig & MT & MT+Orig & Orig & MT & MT+Orig & Orig & MT & MT+Orig &  Orig & MT & MT+Orig \\
\midrule
Dense          &   6 &  3 &      0 &   6 &  3 &      0 &   6 &  3 &      0 &    3 &  1 &      0 \\
Hybrid         &   4 &  2 &      7 &   3 &  2 &      7 &   3 &  2 &      7 &    1 &  2 &      7 \\
Learned-Sparse &   3 &  2 &      0 &   2 &  2 &      0 &   2 &  2 &      0 &    0 &  1 &      0 \\
Sparse         &  15 &  7 &      1 &  15 &  7 &      1 &  15 &  7 &      1 &    2 &  6 &      1 \\
\midrule
Total          &  28 &  14 &     8 &  26 &  14 &     8 &  26 &  14 &     8 &    6 & 10 &     8 \\
\bottomrule
\end{tabular}
\end{table*}

\subsubsection{Overall Retrieval Results}

Despite attracting fewer participants this year,
the submitted runs still include a wide variety of systems.
The CLIR runs summarized in Figure~\ref{fig:clir-ndcg-bar} substantially outperformed the monolingual runs this year;
this is different from last year, when the top runs from CLIR and monolingual tasks had similar nDCG@20. 
The full results are reported in Tables~\ref{tab:zho-full-results}, \ref{tab:fas-full-results}, and \ref{tab:rus-full-results} at the end of the paper. 
The numbers and types of runs clearly indicates that more effort was devoted to CLIR system development this year, and we expect that the monolingual results do not represent the current state-of-the-art for monolingual systems. 
A highlight in the CLIR and MLIR results is the incorporation of the GPT-4 model,
which was used in the top-ranked runs in all the \clirtask\ and MLIR tasks.
Please refer to the system paper~\cite{participants-naverloo} from the \texttt{naverloo} team for more details.%

MLIR is a new task
in NeuCLIR this year, for which the evaluation results are summarized in Figure~\ref{fig:mlir-ndcg-bar}.
The full results are reported in Table~\ref{tab:mlir-full-results} at the end of the paper. 
Among the 24 submitted runs, the majority (18) use the machine-translated documents to create the unified ranked list.
The top-ranked runs for MLIR also used GPT-4;
using out-of-box large language models for reranking proved to be quite effective
(albeit with significantly degraded query-time efficiency).%

\begin{figure*}
    \centering
    \includegraphics[width=\linewidth]{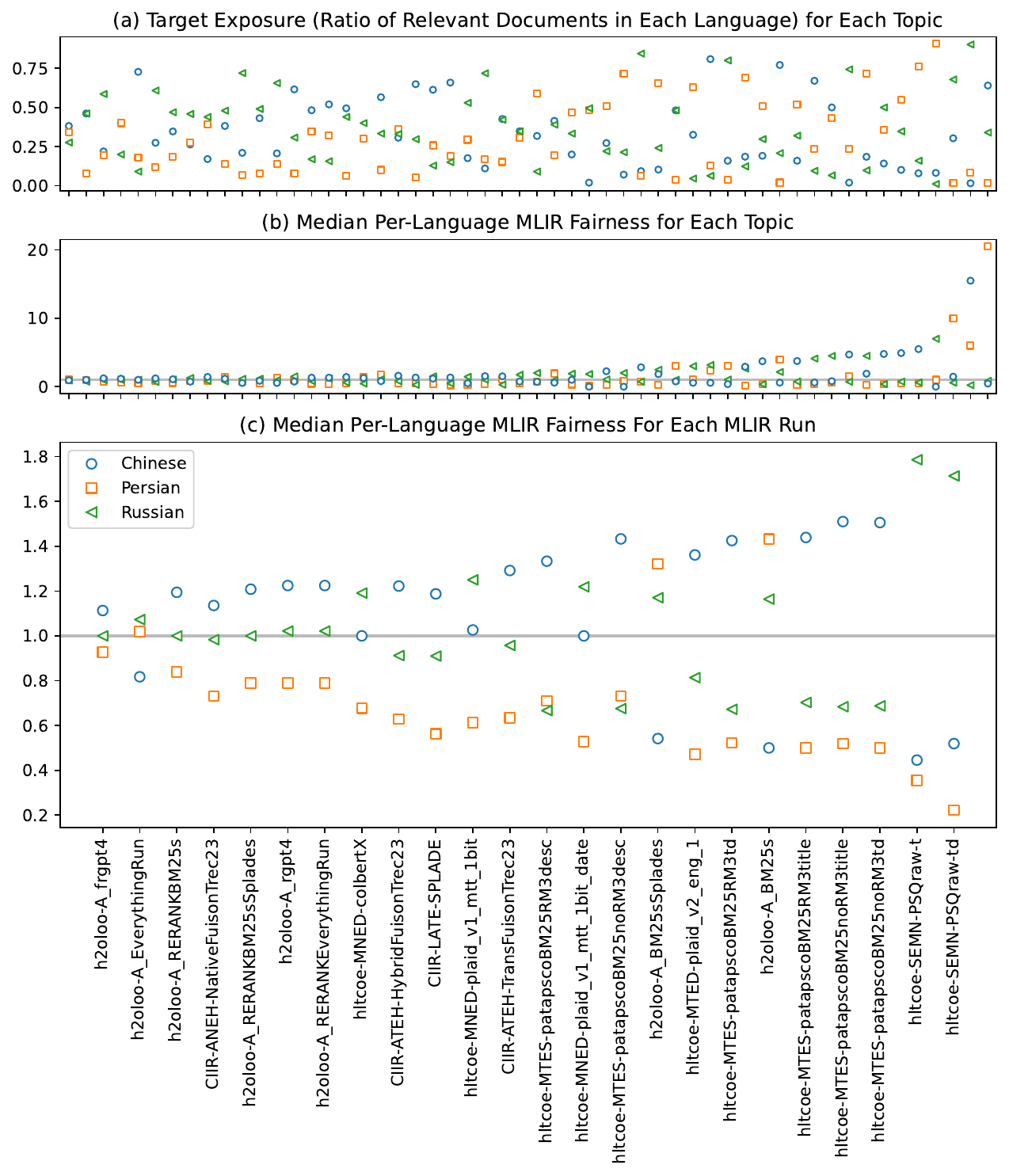}
    \caption{MLIR Target Exposure and Fairness. X-axes of Figures (a) and (b) indicate topics sorted by the range of median per-language fairness among the three languages (range in Figure (b)). The X-axis of Figure (c) indicates runs.}\label{fig:mlir-fairness}
\end{figure*}

\subsubsection{Run Diversity}
Overlap between the top 100 retrieved documents and the retrieved relevant documents of the \clirtask\ tasks
is summarized in Figures~\ref{fig:zho-overlap}, \ref{fig:fas-overlap}, and \ref{fig:rus-overlap} at the end of the paper.
The top-scoring runs are similar to each other in both top-ranked and retrieved relevant documents for all three languages. 
Interestingly, there is a rough correlation between the similarity of the top-scoring runs and the nDCG@20 score,
indicating that the top-scoring runs may be fusing results from different systems and reranking them successfully. 

There are also several clusters of similar runs following the top-scoring group.
Particularly in Persian and Russian, there is a bright triangle at the lower right corner,
which are the BM25 runs contributed by the \texttt{hltcoe} team. 
Among all runs, two particular runs are less similar to others --
\texttt{mContrieverqt} from \texttt{naverloo} and \texttt{blade} from \texttt{umd\_hcil}. 

While the top 100 retrieved documents demonstrate clear clusters,
we do not observe a similar effect among the retrieved relevant documents. 
Even the top-scoring runs all brought different relevant documents into the pool,
which strengthened the reusability of the collection. 

Overlap among the MLIR runs are presented in Figure~\ref{fig:mlir-overlap} (also at the end of the paper).
Similar to the \clirtask\ task, there are clusters of runs that are more similar than others, including the top-scoring group. 
However, these clusters are less apparent, indicating that the methods used in these systems vary.

\begin{figure*}[htbp]
    \centering
    \includegraphics[width=\linewidth]{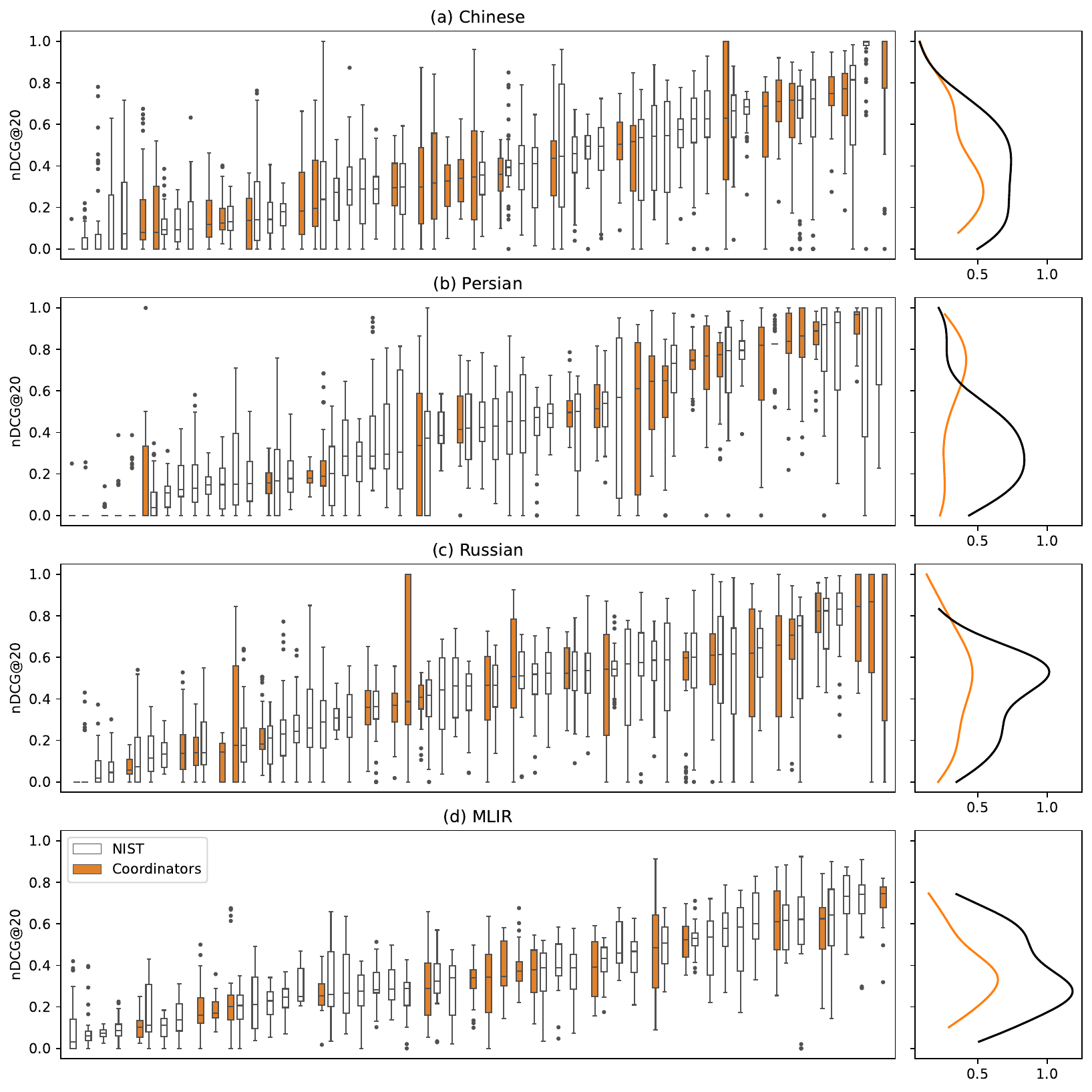}
    \caption{nDCG@20 boxplots (left) and KDE graphs on the median scores (right) of the news task comparing topic developers. Topics are ordered by the median of the scores.}\label{fig:news-boxplot-developer}
\end{figure*}

\subsubsection{MLIR Language Distribution}
Language fairness %
for MLIR is not yet a well-defined concept with widely-reported measures.
We consider one possible measure: 
how the exposure of each language for each topic
matches that language's proportion of relevant documents (which we defined as the target exposure).
This is inspired by the disparate treatment discussed in \citet{singh2018fairness} and the R@MLIR-Relevant measure proposed by \citet{lawrie2022multilingual}. 
Specifically, we define per-language document exposure %
as the proportion of documents in a language at or above
rank $R$,
regardless of relevance, where $R$ is the total number of relevant documents for the topic across all languages. 
Similarly, we define the target exposure
as the proportion of the relevant documents for a language at or above rank $R$, divided by $R$, which are plotted in Figure~\ref{fig:mlir-fairness}(a). 

\begin{figure*}[hbtp]
    \centering
    \includegraphics[width=\linewidth]{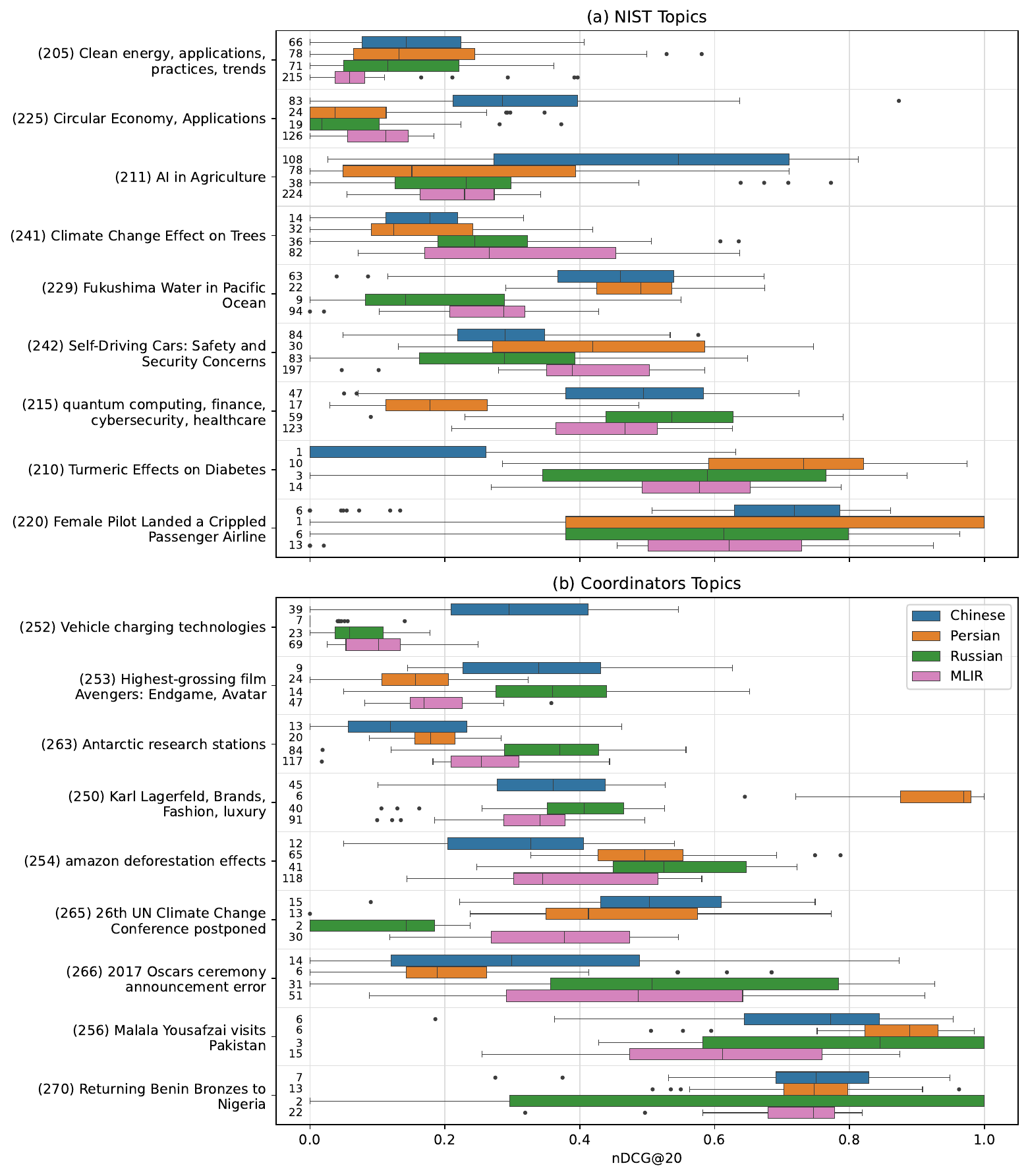}
    \caption{Sampled nDCG@20 boxplot of the news task comparing among the \clirtask\ and MLIR tasks. The numbers at the left of each boxplot are the numbers of relevant documents in the corresponding task and topic. }\label{fig:news-boxplot-sampled}
\end{figure*}

\begin{figure*}[ht]
    \begin{subfigure}[t]{\textwidth}
        \centering
        \caption{Leave-one-run-out}\label{fig:news-loro}
        \includegraphics[width=\linewidth]{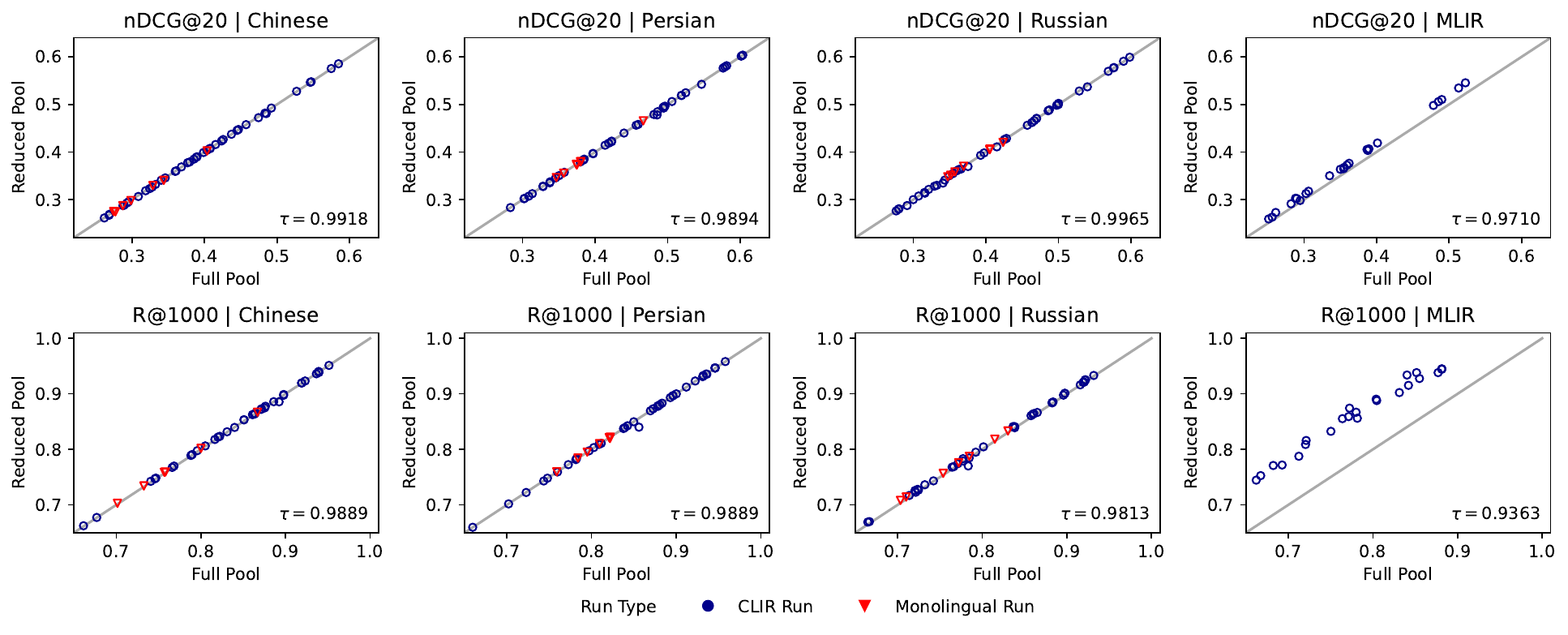}
    \end{subfigure}
    \begin{subfigure}[t]{\textwidth}
        \centering
        \caption{Leave-one-team-out}\label{fig:news-loto}
        \includegraphics[width=\linewidth]{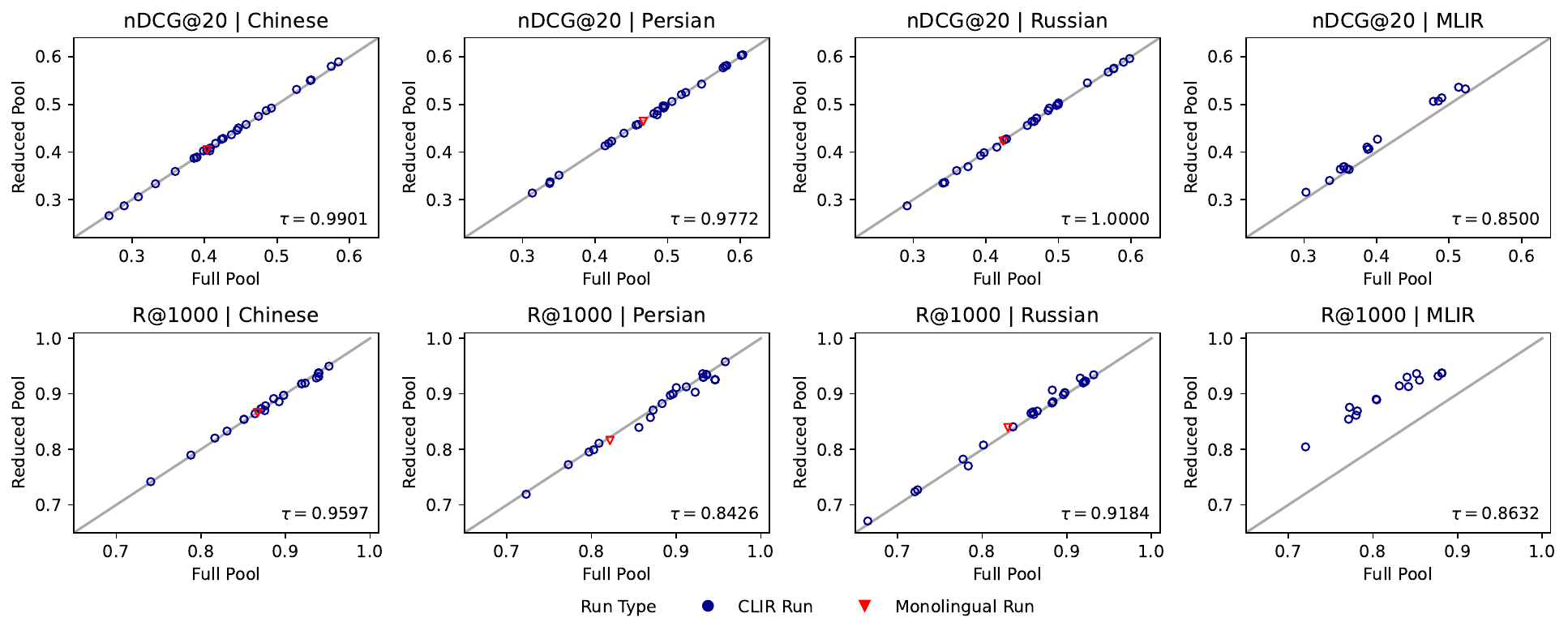}
    \end{subfigure}
    \caption{Collection Reusability Study with leave out experiments. Leave-out experiments on MLIR are constructed only using MLIR runs, which assumes only the MLIR task was held. Since there are 71 runs serving only the purpose of pool enrichment submitted by the Coordinators under the team name \texttt{hltcoe}, we only conduct experiments with actual participating runs without the enrichment ones in this study.}
\end{figure*}

\begin{figure*}
    \centering
    \includegraphics[width=\linewidth]{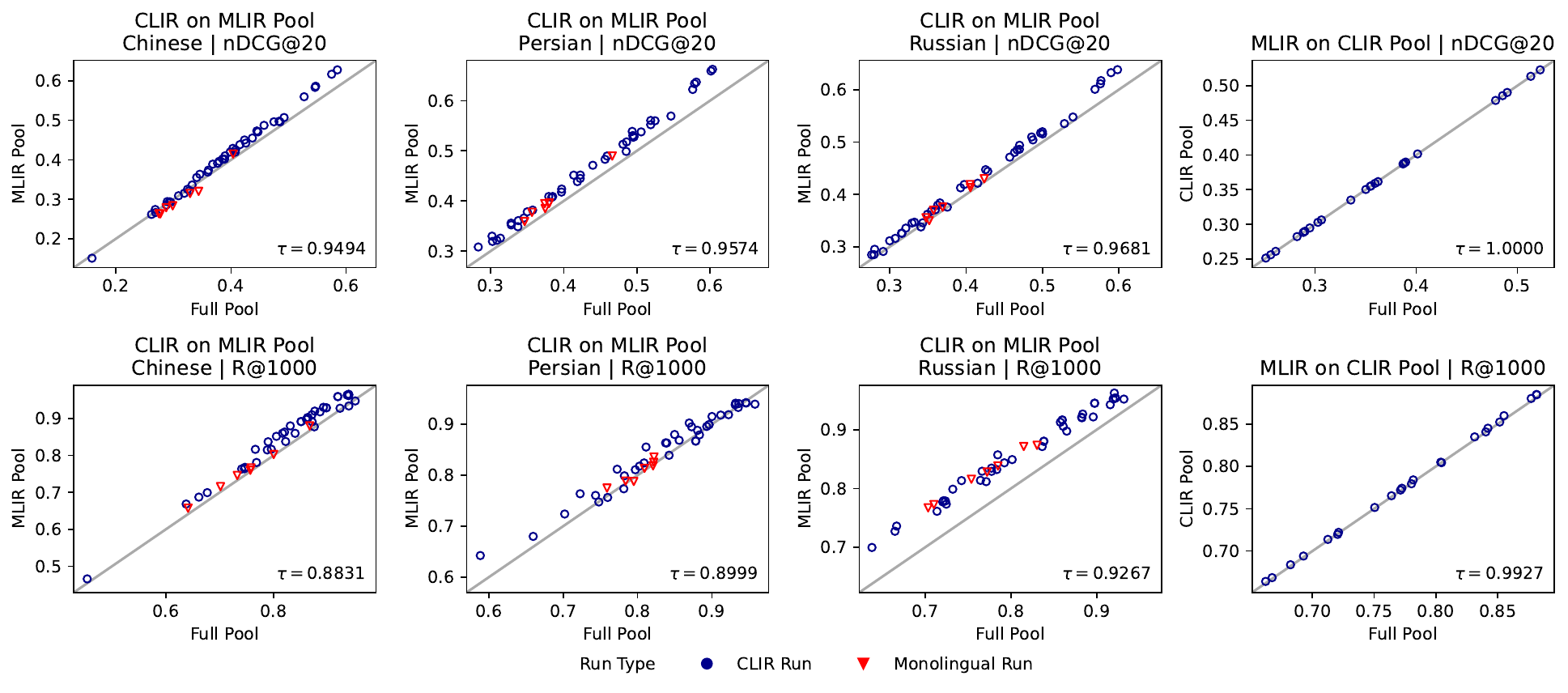}
    \caption{CLIR runs evaluated on MLIR-only Pool and vice versa.}\label{fig:news-clir-mlir-pool-on-each-other}
\end{figure*}

We report the ratio between the per-language document exposure to the target exposure, which we refer to as the per-language \textit{MLIR fairness} for convenience in this report. 
Figure~\ref{fig:mlir-fairness}(b) and (c) report the median MLIR fairness among the submitted MLIR runs for each topic and for each run, respectively. 
Topics in Figures~\ref{fig:mlir-fairness}(a) and (b) are sorted by the range of the median per-language fairness among the three languages, which is an indicator of how languages are treated differently among the submitted runs for a given topic.
Runs in Figure~\ref{fig:mlir-fairness}(c) are sorted similarly, but the medians were calculated among all topics. \
We use the range of medians as a proxy for the overall fairness of the topic or the run. However, we plot the per-language fairness for better illustration and analysis. 
The intuition behind this ratio is that the closer to 1.0,
the closer a system (or for Figure~\ref{fig:mlir-fairness}(b) the set of all systems for a topic) comes to presenting each language based on its utility,
which is the proportion of relevant documents it carries.

It is interesting to observe that the ordering of the effectiveness among the MLIR runs is not the same as the ordering of the MLIR fairness (Kendall's $\tau$ = 0.61). 
While there is still a positive correlation between the two, this analysis suggests that the most effective systems may not be the fairest systems for the three languages, indicating room for advancement in MLIR retrieval systems. 

Among the topics, some topics receive particularly unfair treatment among the systems.
These topics generally have an imbalance in the number of relevant documents among the three languages, which may make it harder for systems to provide fair effectiveness for all languages. 
While this distribution discrepancy among the topics may not be undesirable, 
such phenomenon still requires more investigation into their development processes

\subsubsection{Topic Difficulty}

Figure~\ref{fig:news-boxplot-developer} illustrates the nDCG@20 score distribution among the submitted runs in all four tasks
using the news collection. 
Topics for \clirtask\ tasks demonstrate a wide spectrum of difficulties. 
However, some runs still perform well on difficult topics (outlier dots in the boxplots), 
while some runs also performed poorly on relatively easy topics. 

For MLIR, the easiest topics received a maximum of 0.81 nDCG@20
compared to the maximum of 1.0 in the \clirtask\ tasks.  
or that the systems submitted to the MLIR task were less developed. 
Nevertheless, the MLIR topics also demonstrate a wide spectrum of difficulty,
ranging from a median of lower than 0.1 to higher than 0.8. 

Since NIST and the Track Coordinators used different methods to develop topics,
it is useful to investigate whether the topics in each set
(as described in Section~\ref{sec:clir:rel-judgment})
demonstrate distinct characteristics. 
Colored boxplots in Figure~\ref{fig:news-boxplot-developer} indicate Coordinator topics. 
Despite demonstrating different extreme characteristics, the topic sets are similar. 
The Coordinator topics spread across the entire spectrum and do not have a clear cluster. 
Plotting the median scores of the topics in KDE plots
(right graphs of Figure~\ref{fig:news-boxplot-developer}), 
topics developed by NIST and by the Coordinators have similar distributions
except for differences in peak densities, which result from the differences in the number of topics. 

Figure~\ref{fig:news-boxplot-sampled} demonstrates a sample of topics grouped by topic set (NIST or Coordinator topics). 
Among NIST topics, several, e.g., Topics 229, 215, and 210,
are particularly easy for two languages and difficult for the third. 
This phenomenon is likely due to the number of relevant documents contained in each language,
which in turn could be an artifact of topic development in language pairs. 
However, Coordinator topics demonstrate a similar trend
despite being developed using machine-translated documents for all languages;
this indicates that it might not be feasible to identify topics that are equally attested in all three languages
without having language skills in all languages.

\subsubsection{Collection Reusability}

\begin{figure*}[t]
    \begin{subfigure}[t]{\textwidth}
        \centering
        \caption{Leave-one-run-out}\label{fig:news-loro-participant}
        \includegraphics[width=\linewidth]{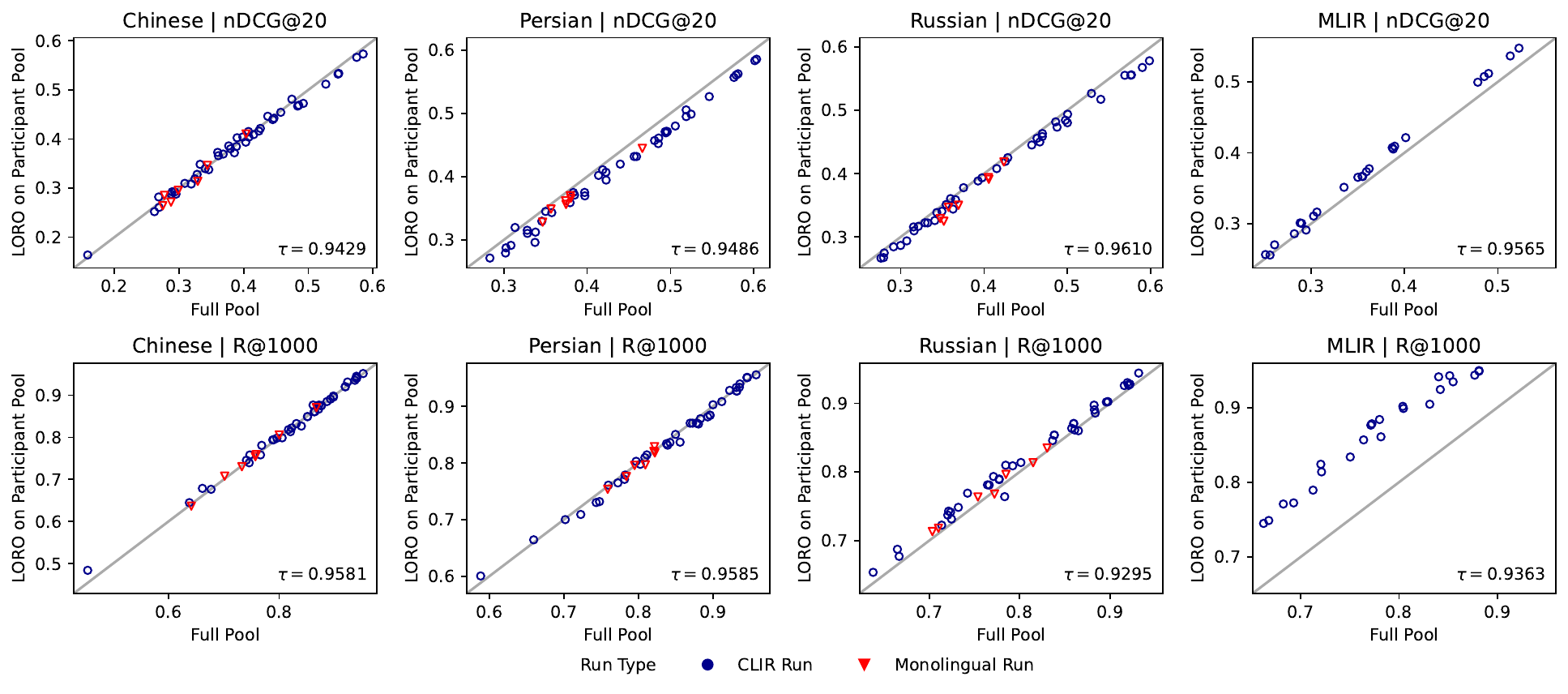}
        
    \end{subfigure}
    \begin{subfigure}[t]{\textwidth}
        \centering
        \caption{Leave-one-team-out}\label{fig:news-loto-participant}
        \includegraphics[width=\linewidth]{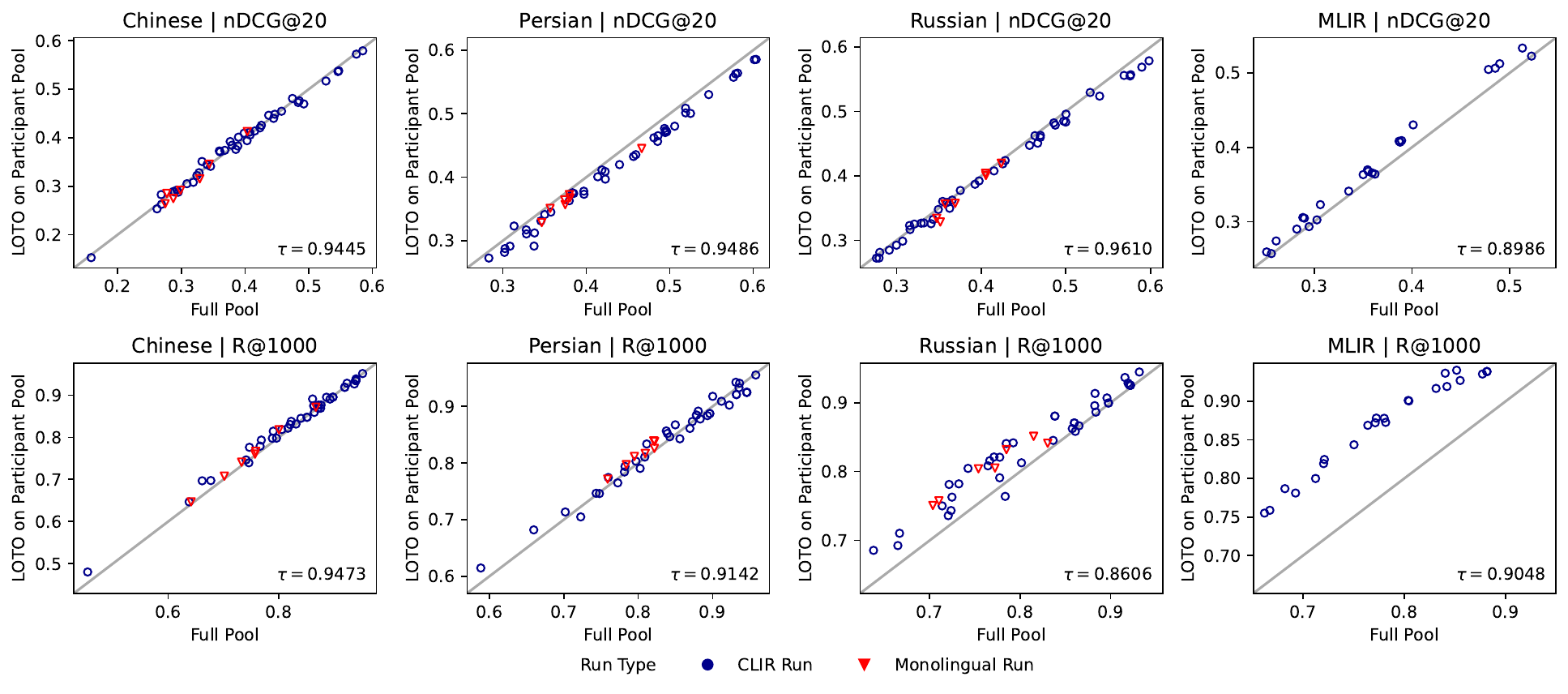}
    \end{subfigure}
    \caption{Collection Reusability Study without the additional 71 Coordinator's runs, which are all baseline models for enriching the pool. Leave-out experiments on MLIR are constructed only using MLIR runs, which assumes only the MLIR task was held. }
\end{figure*}

To test the reusability of the collections created,
we conducted leave-one-run-out (LORO) and leave-one-team-out (LOTO) experiments;
their results are summarized in Figures~\ref{fig:news-loro} and \ref{fig:news-loto} respectively. 
In these experiments, the top-ranked documents 
(either 20 or 50 depending on the type of run)
that contributed to the full pool from each run (LORO) or team's runs (LOTO)
are removed from the relevance judgment pool if those same documents were contributed to the pool by no other run (LORO) or team (LOTO);
this creates a reduced pool that simulates future runs that do not contribute to the pools. 
We use the reduced pool to evaluate the removed run
and compare the scores with those calculated based on the full pool. 
Note that a top-ranked document from the removed run may remain in the pool
since it could be included in another run.
However, since runs submitted by the same team share more similarity than others,
LOTO removes more documents from the pool,
resulting in more volatile results~\cite{lawrie2022overview}.

\clirtask\ tasks are relatively stable at the top of the ranking when measuring nDCG@20 in both LORO and LOTO experiments,
indicating that the collection can evaluate runs and systems that are not included in the pooling without much bias. 
However, when evaluating recall at 1000,
LOTO experiments indicate that a set of systems that are very different from the ones in the pools
may discover more relevant documents. 
Since runs from each team tend to discover some unique relevant documents,
removing all runs from a team results in fewer relevant documents in each topic,
which leads to inflated recall values. 
However, such volatility is also likely to be a result of having fewer participating teams in the track this year.  

Creating pools solely based on MLIR runs seems suboptimal.
The four figures at the right-most column in Figure~\ref{fig:news-loro} and Figure~\ref{fig:news-loto}
are LORO and LOTO experiments only using the MLIR runs without the contribution of \clirtask\ runs. 
Since we have fewer submitted runs in the MLIR task, both LORO and LOTO result in scores that are more volatile. 
Particularly for LOTO, the effect of finding unique relevant documents is also larger
(as seen in Figure~\ref{fig:mlir-overlap}),
resulting in even larger variation when leaving one team out of the pools. 
Using the same pool depth for the MLIR and the \clirtask runs may also contribute to this problem as the combined pool size from \clirtask is larger than the ones built from the MLIR tasks.
Therefore, pooling only the MLIR runs results in a less reusable collection than does pooling on both \clirtask\ and MLIR runs. 

We ask a further question: what if we only host the \clirtask\ task but not the MLIR task? 
Figure~\ref{fig:news-clir-mlir-pool-on-each-other} summarizes the experiments
constructing pools with only \clirtask\ or only MLIR runs and evaluating on the other set of tasks. 
Evaluating \clirtask\ tasks with qrels pooled from the MLIR runs
(left three columns) indicates a similar trend in Figure~\ref{fig:news-loro} and Figure~\ref{fig:news-loto},
indicating that the reusability is more concerning than when pooling on all tasks. 

However, using qrels produced by the \clirtask\ pool is more feasible. 
In the right-most columns of Figure~\ref{fig:news-clir-mlir-pool-on-each-other},
the nDCG@20 scores produced with the \clirtask\ pool are identical to those calculated from the full pool,
which assures us that all top-ranked relevant documents retrieved by the MLIR systems were also retrieved by the \clirtask\ systems
and included in the pool. 
While recall at 1000 is not perfectly correlated between the two versions,
they are extremely close with Kendall's $\tau$ of 0.99. 
The stable evaluation results using only the \clirtask\ pool on the MLIR runs
suggest that NeuCLIR 2022 qrels can reasonably be used to evaluate MLIR runs as well.\footnote{There are 41 trilingual topics in NeuCLIR 2022 that can be used to evaluate MLIR systems by combining the three collections and qrels together to form the MLIR collection and qrels. } 

Since the number of participating teams is fewer than last year,
the Coordinators enriched the pools
by adding additional runs under the team name of \texttt{hltcoe}). 
In total, we added 71 submissions across the three \clirtask\ tasks and the MLIR task. 
To verify whether this pool enrichment is beneficial,
we experimented with removing them from the pools and tested for reusability.
The results are summarized in Figures~\ref{fig:news-loro-participant} and \ref{fig:news-loto-participant}. 
The results indicate that both LORO and LOTO become less stable
and demonstrate a weaker correlation between the scores calculated from the full (including enrichment runs) and reduced pool (excluding enrichment runs) in almost all tasks and metrics. 
Nevertheless, we can still conclude that including additional pool enrichment runs provides extra stability to the final collection. 

%% file: tech-2-task.tex
\input{_fig_example_csl_doc}
The Technical Documents Task is a new pilot task this year.
It is a cross-language ad hoc retrieval task, with English queries and Chinese documents.
The key distinguishing feature of this task is the technical nature of the documents.
The document collection consists of almost 400,000 abstracts of Chinese academic papers and theses,
drawn from domains such as Chemistry, Physics, and Computer Science.
Figure~\ref{fig:techdoc} shows a sample document from the collection,
along with a machine translation of the document into English.
While the pilot task this year used a small number of topics,
we expect a full task next year to allow researchers to gauge the effectiveness of existing CLIR approaches on technical documents,
and to identify along which dimensions those systems need improvement.

This task contains the same three settings as the newswire \clirtask\ task,
namely ad hoc CLIR, reranking CLIR, and monolingual retrieval.
To support reranking, teams were given the output of a BM25 retrieval system,
which used document translation to cross the language barrier. 
The ID for that run is \texttt{tech-hltcoe-MTES-} \texttt{patapsco\_bm25\_td\_rm3}. 

Participants in the Chinese news \clirtask\ task were asked to also submit one or more baseline technical document runs
using their news retrieval system.
In addition to English topics, human %
translations of topics into Chinese were made available to participants,
who were also encouraged to submit monolingual runs using those topics.

%% file: _fig_example_csl_doc.tex
\begin{figure*}
    \centering
    \begin{tabular}{l|p{2.5in}|p{2.5in}}
    \toprule
    Key     &   Original  & Google Translation \\
    \midrule
    doc\_id & csl-121374 & csl-121374 \\
    Title & \begin{CJK*}{UTF8}{gbsn}(S)-2-氨基-1,1-二苯基-1-丙醇的合成和外消旋Droxidopa前体的拆分\end{CJK*}
    & (S) -2-amino-1,1-two phenyl-1-propane synthesis and external anti-rotation DroxIDOPA foretophylline is split \\
    Abstract &
    \begin{CJK*}{UTF8}{gbsn}(S)-2-氨基-1,1-二苯基-1-丙醇是一种合成多种手性助剂的重要中间体,也用于外消旋屈昔多巴前体化合物的拆分.从价廉易得的L-丙氨酸出发,通过四步反应制得,总收率55.6\%L-丙氨酸经甲酯化,苄氧羰基保护制得的L-2-苄羰基氨基丙酸甲酯与苯基溴化镁反应制得(S)-2-苄氧羰基氨基-1,1-二苯基-1-丙醇.是在5\%Pd/C催化加氢下脱除苄氧羰基得到标题化合物.该制备方法涉及的中间体及目标化合物易于纯化,总收率高且重现性好.我们用制得的氨基醇能成功地拆分外消旋苏式屈昔多巴前体化合物3-3,4-二苄氧苯基)-N-苄氧羰基丙氨酸.\end{CJK*}
    & \begin{CJK*}{UTF8}{gbsn}(S) -2-amino-1,1-two phenyl-1-propionol is an important intermediate of a synthetic multi-hand assistant. It is also used for demolition Divided. From the price of low-priced L-Alanine, obtained through four-step reaction system, the total revenue is 55.6\%L-alanine through methyl, and the L-2-cymbal base protected by the oxygen oxygenyl The reaction of aminopenate metropolis and phenyl bromide reaction (s) -2-苄 oxygenyl amin amino-1,1-two phenyl-1-propyol. The divisor oxygen cymbal group gets the title compound. The intermediate and target compounds involved in the preparation method are easy to purify, the total yield is high and the reproducibility is good. Daba pre-body compound 3-3,4-two-苄 oxygenyl) - 苄 oxygenyl alanine. \end{CJK*}\\
    Keywords & \begin{CJK*}{UTF8}{gbsn}拆分剂, (S)-2-胺基-1,1-二苯基-1-丙醇, β-氨基醇, 屈昔多巴\end{CJK*}

    & Disassembly, (S) -2-aminel-1,1-two phenyl-1-1-propylene, $\beta$-amino alcohol, Koshidaba \\
    Category & \begin{CJK*}{UTF8}{gbsn}工学\end{CJK*} & Engineering \\
    Discipline & \begin{CJK*}{UTF8}{gbsn}化学/化学工程与技术\end{CJK*} & Chemistry and Chemical Engineering \\
    \bottomrule
    \end{tabular}

    \caption{Example document from the CSL dataset.}
    \label{fig:techdoc}
\end{figure*}

%% file: tech-3-docs.tex
\subsection{Documents}

The documents for this task were abstracts from the Chinese Scientific Literature (CSL) dataset~\cite{li-etal-2022-csl}.
The dataset contains 396,209 journal abstracts from 1,980 academic Chinese journals spanning 67 general disciplines,
where Engineering, Science, Agriculture, and Medicine dominate.
Articles were published between 2010 and 2020.
The abstracts were originally obtained from the National Engineering Research Center for Science and Technology Resources Sharing Service (NSTR).\footnote{https://nstr.escience.net.cn/}
For use in this pilot task, we reorganized the dataset into the same format as the news documents. 

%% file: tech-4-topic.tex
\subsection{Topics}

Topic creation was accomplished by seven graduate students 
in biology, biomedical engineering, chemistry, chemistry and biomolecular engineering, economics, and physics %
at The Johns Hopkins University.
Annotators were hired based on their Chinese language skills and their familiarity with scientific research.
During an interview, students were asked to describe their research area in both Chinese and English.
They then were asked to choose a research topic they were familiar with,
enter a Chinese language query on that topic into an interactive search system that returned documents from the CSL dataset,
and read and briefly summarize the top returned documents to determine whether they were relevant to their search.
The purpose of this part of the interview was to ensure that the collection contained documents related to their area of research.
Of the seven students, one was a Ph.D. student and the others were Masters students.
Three identified as women and four as men.%

Once hired, each annotator participated in a three hour online training session.
During the training, the topic creation task was explained.
Then each person worked independently to create their first topic.
During that process, two of the Coordinators reviewed their ongoing work.
This time was used to ensure that topics had a suitable level of specificity,
and that the tool was being used properly to determine whether abstracts on the topic existed in the collection.
After the training, assessors were asked to spend up to a total of ten hours creating four to six topics.
Five assessors created six topics apiece, while two assessors created five topics,
yielding a total of forty topics for the pilot task.

The English %
title, description, and narratives were reviewed by a Coordinator
to ensure that the topic was sufficiently descriptive.
The topic %
was also checked for grammar and spell-checked.
In some cases, the assessors were asked to revise topics that appeared to be too vague or were not understandable. After any revision, assessors checked the translation to ensure that it incorporated any changes. The translations did not undergo any external quality control.
In the end forty topics were distributed to participants. %

%% file: tech-5-judgment.tex
\subsection{Relevance Judgments}

\begin{table*}[thp]
\caption{Relevance judgments for technical document task.}\label{tab:tech-judge}
    \centering

\begin{tabular}{l|ccccc|c}
\toprule
                                  &  Chemistry &  Economics &  Physics &  Biology &  Medicine &  Overall \\
\midrule
\# Topics Judged                  &          4 &          6 &        7 &        9 &        13 &       39 \\
\midrule
Avg. \# Judgments / Topic         &     363.25 &     239.17 &   241.57 &   346.56 &    276.38 &   289.51 \\
\midrule
Avg. \# Somewhat Valuable / Topic &       9.75 &      18.67 &    11.33 &     7.78 &     10.00 &    11.03 \\
Avg. \#  Very Valuable / Topic    &       6.50 &       3.80 &     9.50 &     8.62 &      5.83 &    6.91  \\
\bottomrule
\end{tabular}

\end{table*}

Assessors then participated in a second online training session, lasting one hour, that focused on the relevance judgment process. In addition, instructions written in English were provided for completing relevance judgments.
Relevance for the Technical Documents pilot task differed somewhat from the usual TREC view of relevance.
Assessors were asked to imagine that they were writing the background section or the related work section
of a scientific paper on the topic they had created.
They were asked to evaluate whether they would plan to read the paper being judged based on its abstract
so as to possibly cite the paper in their related work section.

They answered two questions about each document:
\begin{definition}
 Does this document contain central information?
 \begin{description}
 \item
 [Yes] There is information in the abstract related to their search topic. 
 \item [No] 
There is no information in the abstract related to their search topic. 
\item [Unable to judge] The document was not viewable in the document viewer panel.
\end{description}
\end{definition}
\begin{definition}

How valuable is the most important information in this document?
 \begin{description}
 \item
[Very Valuable] One would definitely read the paper associated with this abstract when writing the related work section for this research topic.
\item
[Somewhat Valuable] If one had enough time one would read the paper,
because it might have something that could appear in the related work section,
but confidence about that is low.  
\item
[Not that Valuable] One is unlikely to read the paper
because one does not expect to find in it information that one would cite in the related work section. 
\end{description}
\end{definition}

Assessor progress was tracked;
when very few or an excessive number of relevant documents were found, or too little time was spent completing the task,
they were asked to rejudge the pool.
In one case, an assessor judged their pools twice and had so little consistency
(only one document was judged relevant twice) that
all judgments from that assessor were discarded.
Fortunately, another assessor was sufficiently familiar with the area of science represented by that removed assessor's topics
to judge five of the six topics.
All but one assessor was able to complete relevance assessment in ten hours or less;
the outlier assessor required twelve hours to complete the task. 

Pools were created from the top twenty documents of each submitted run.
In the end thirty-nine topics were used to judge system performance.
Table~\ref{tab:tech-judge} contains information concerning the judgments and the average number of very valuable and average number of somewhat valuable per topic in the judged pools. %

%% file: tech-6-resource.tex
\subsection{Additional Resources}

In addition to the document collection itself and the resources already described in~\ref{sec:clir-resources},
the track provided translations into Chinese of the topic fields,
and translations into English of the document texts.
These translations were obtained from the online \textit{Google Translate} service.

%% file: tech-7-participation.tex
\subsection{Participation}

The Technical Documents pilot task had five participants with 51 runs:
\begin{itemize}
    \item Allen Institute for AI
    \item Johns Hopkins University HLTCOE
    \item University of Massachusetts CIIR
    \item University of Southern California ISI
    \item University of Waterloo and Naverlab
\end{itemize}
Four of these teams used their Chinese news systems; the system from the Allen Institute for AI was originally designed for technical documents, but not tuned specifically for the CLIR task. %

%% file: tech-8-results.tex
\subsection{Analysis of Official Submissions}

\begin{table*}
\caption{Technical Document pilot task CLIR and monolingual runs, with and without document machine translation.}\label{tab:tech-agg}
    \centering

\begin{tabular}{llc|ccc|ccc|ccc}
\toprule
\multirow{2}{*}{Setting}  & \multirow{2}{*}{Indexed Docs}   & \multirow{2}{*}{\# Runs} &
                        \multicolumn{3}{c|}{nDCG@20} & \multicolumn{3}{c|}{MAP} &\multicolumn{3}{c}{R@1000}\\
&                 &      &   Mean  & Median &   Max  &   Mean  & Median &   Max &   Mean  & Median &   Max \\
\midrule
\multirow{3}{*}{CLIR}
& Original Docs   &   23 &   0.301 &  0.311 &  0.394 &  0.206 &  0.206 &  0.287 &  0.733 &  0.755 &  0.865 \\
& Translated Docs &   10 &   0.270 &  0.283 &  0.341 &  0.180 &  0.186 &  0.235 &  0.743 &  0.763 &  0.839 \\
& Both            &    9 &   0.456 &  0.484 &  0.496 &  0.351 &  0.378 &  0.391 &  0.917 &  0.925 &  0.944 \\
\midrule
\multirow{2}{*}{Monolingual}
& Original Docs   &    6 &   0.351 &  0.355 &  0.410 &  0.264 &  0.273 &  0.308 &  0.773 &  0.785 &  0.813 \\
& Translated Docs &    3 &   0.234 &  0.237 &  0.240 &  0.165 &  0.164 &  0.171 &  0.797 &  0.799 &  0.815 \\
\bottomrule
\end{tabular}

\end{table*}

\begin{figure*}
    \centering
    \includegraphics[width=\linewidth]{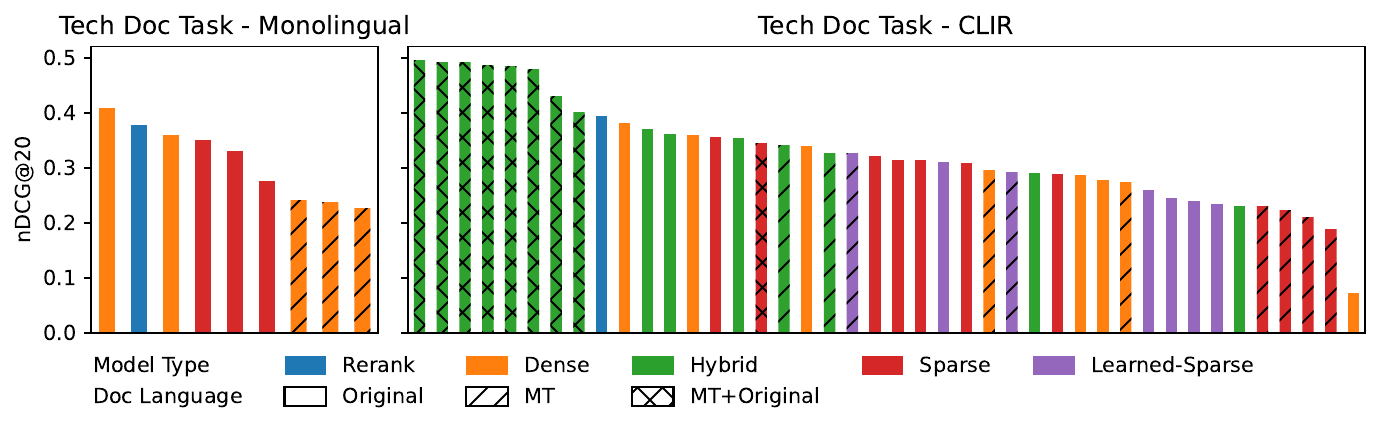}
    \caption{Technical Document pilot task nDCG@20.}\label{fig:tech-ndcg-bar}
\end{figure*}

The Technical Document pilot task challenges participants to develop more effective CLIR models
when documents may be difficult for available MT models to translate.
Figure~\ref{fig:tech-ndcg-bar} summarizes the runs submitted for the task. 
Please refer to Table~\ref{tab:tech-full-results} at the end of this paper for the full evaluation results. 
The \texttt{naverloo} team also contributed the top-scoring runs to the Technical Document task
with their ensemble of systems approach.

\begin{figure*}
    \centering
    \includegraphics[width=\linewidth]{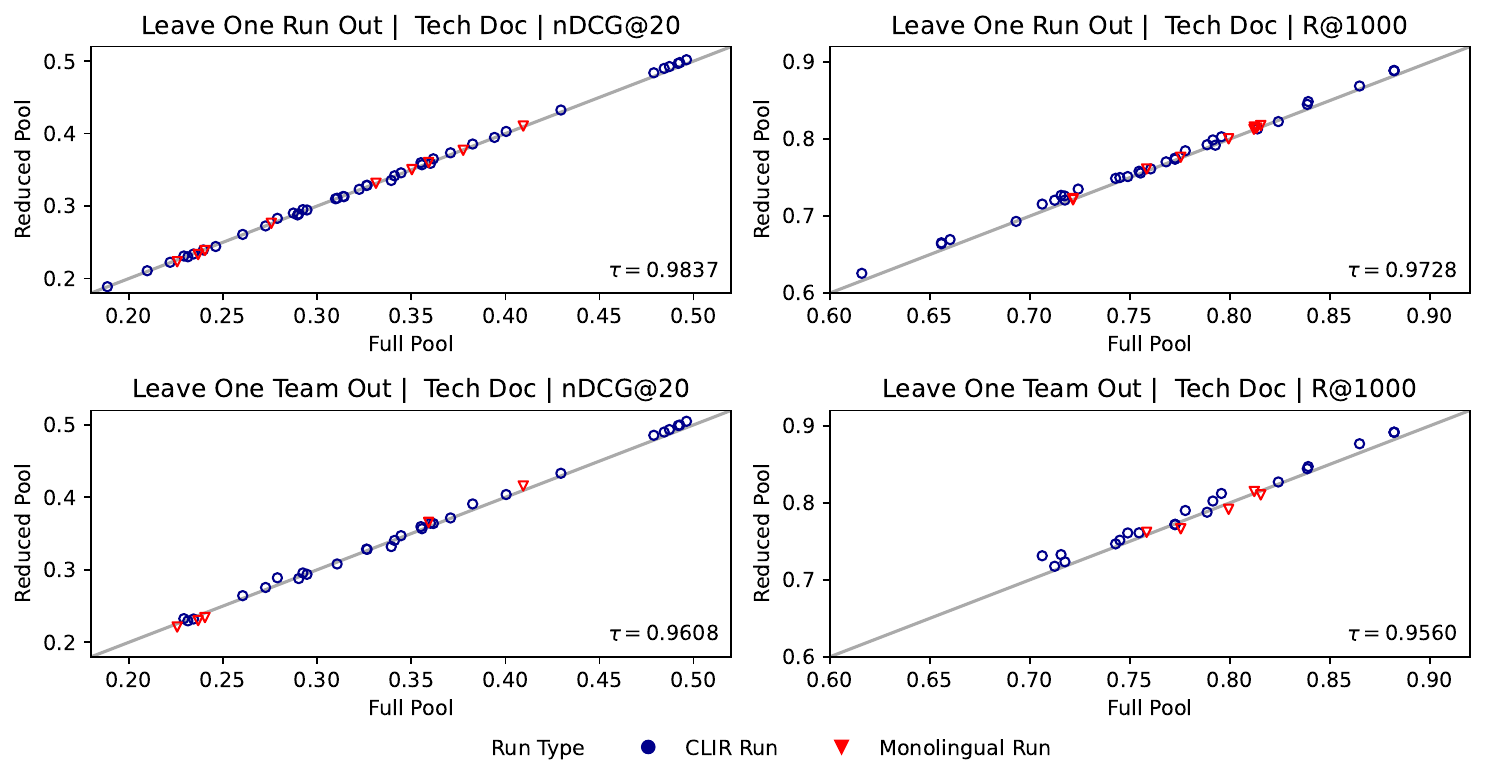}
    \caption{Leave One Run Out and Leave One Team Out Experiments on the Technical Document collection. Since there are 16 runs submitted by Coordinators recorded under the team \texttt{hltcoe} serving only for pool enrichment, we exclude them from leave-one-team-out experiments for the team \texttt{hltcoe}.  }\label{fig:tech-loo}
\end{figure*}

Since we expect that machine translation models were not tuned for the technical vocabulary in the documents,
we hypothesized that systems using machine-translated documents would be less effective
than models directly operating on documents in their native form.
Table~\ref{tab:tech-agg} reports aggregated metrics over \clirtask and monolingual 
 runs using original or machine-translated documents, or both. 
While runs only using document translation generally perform worse than the runs using the original documents in both \clirtask and monolingual settings,
the two seem to provide complementary information and, thus, foster improved effectiveness when using both (only \clirtask submissions using both). 
However, results presented in Table~\ref{tab:tech-agg} are aggregated over different systems with different designs. 
More investigation is needed to determine whether using both original and machine translated documents
leads to more effective retrieval results when holding other attributes of the system design constant. 

\subsubsection{Run Diversity}
As shown in Figure~\ref{fig:tech-overlap} at the end of this paper,
the top-scoring runs retrieve similar sets of top-100 documents. 
Similar to the News document task,
the top-scoring runs are more similar to each other than to the other runs. 
However, we do not observe the clusters of similar runs that we did in the News document task,
except for the top-performing runs. 
There are still pairs or groups of runs that retrieve similar top-100 documents,
but with relatively larger differences in nDCG@20.

\subsubsection{Collection Reusability}

We conducted Leave-One-Run-Out and Leave-One-Team-Out experiments using the Technical Document pilot task runs. 
The Leave One Run Out (LORO) results, 
where each run is evaluated with the qrels from which its contribution was removed
(as if it were a future run), are shown in the top two graphs in Figure~\ref{fig:tech-loo}.
Those results are similar to the results obtained using the full pools;
this would suggest that future runs can be evaluated fairly with the existing qrels. 

However, the Leave One Team Out (LOTO) results shown in the bottom two graphs in Figure~\ref{fig:tech-loo} (
leaving out all runs contributed by a team to the qrels and then evaluating all their runs)
resulted in larger metric value shifts, particularly for R@1000.
We expect that the lower participation in this task may explain some of the larger variation observed here.
However, the correlation is still high (Kendall's $\tau$ > 0.95), which indicates that the future systems are still can still be fairly compared with this collection.

\subsubsection{Topic Difficulty}
\begin{figure*}
    \centering
    \includegraphics[width=\linewidth]{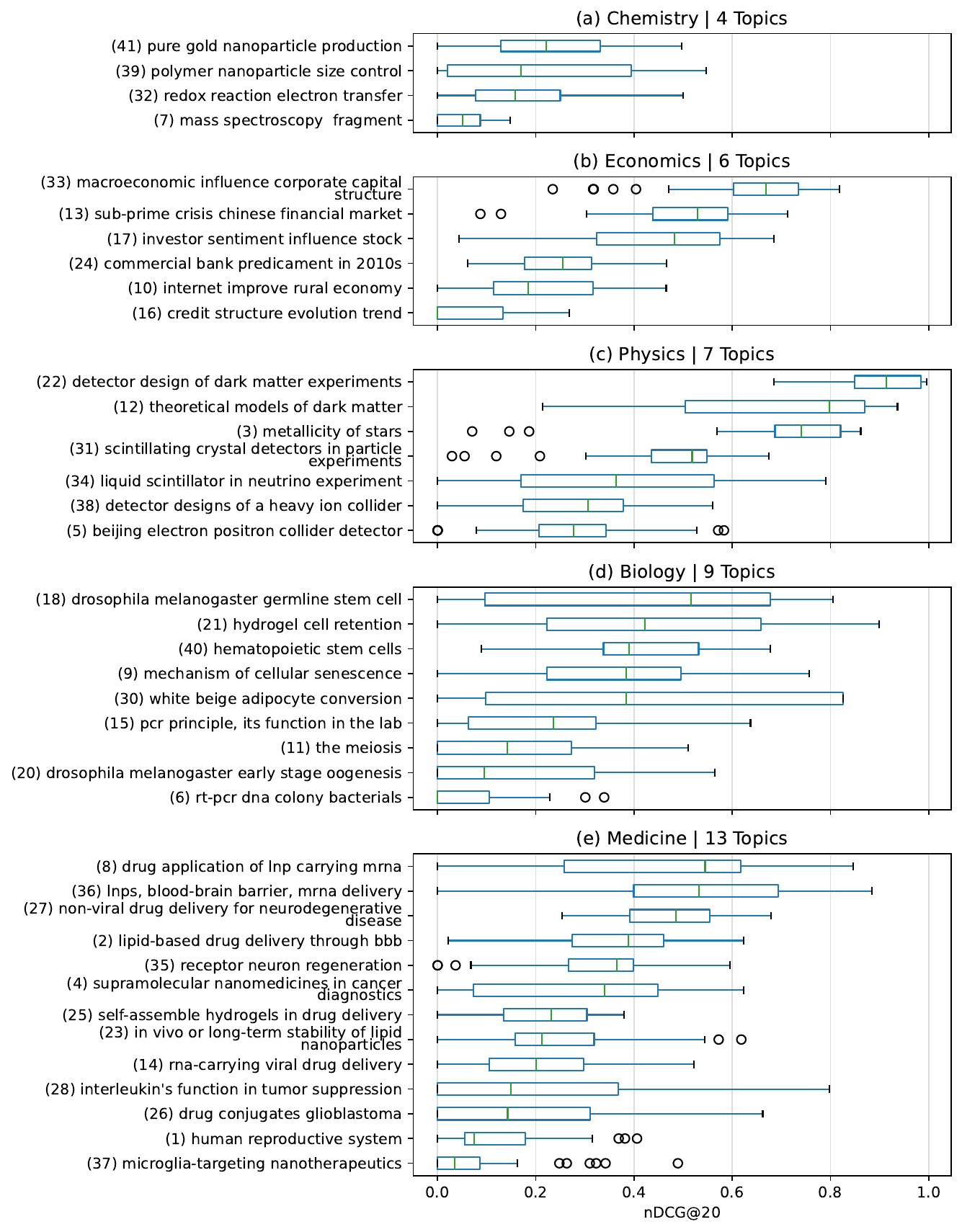}
    \caption{nDCG@20 Boxplot for sampled topics in the Technical Document task.}\label{fig:tech-boxplot}
\end{figure*}

Finally, we demonstrate the per-topic performance distribution across all submitted runs. 
Summarized in Figure~\ref{fig:tech-boxplot},
39 topics are categorized into five disciplines.
Medicine is the largest group.
This is a result of recruiting graduate students from The Johns Hopkins University,
which is notably strong in medicine, as topic developers. 

Topic difficulty varies, but the majority of topics (27) have a median nDCG@20 below 0.4. 
There are some particularly hard topics, such as Topics 7, 16, and 37.
However, some runs still achieved notably higher effectiveness on these difficult topics,
indicating that the relevant documents they found are not unretrievable.

%% file: 9-future.tex
\section{Future Track Directions}

The NeuCLIR track will run for a third year at TREC 2024.
In 2024, we plan to capitalize on successful aspects of the 2023 edition,
and to revise the track timeline to encourage additional participation.

\subsection{What's New in NeuCLIR 2024?}

First, we will promote the Chinese Technical Document task from a pilot task to a full task. 
Concretely, NeuCLIR 2024 plans to produce a CLIR test collection for technical documents with at least fifty topics.
As we have noited, we would expect general purpose machine translation systems can have difficulty translating technical documents,
due to the specialized terminology they use.
A full CLIR task to retrieve such documents will help to ascertain
the degree to which CLIR techniques 
can address this challenge. 
We will use the same collection of Chinese dissertation abstracts from multiple disciplines
that we used in the NeuCLIR 2023 Technical Docuemnts pilot task as the document collection for this task. 
The key challenge will be finding annotators with both the language skills and the technical experience needed
to produce representative topics and accurate relevance judgments. 
The 2024 full task will expand from our initial focus on graduate students at one university to include a more diverse group of annotators from additional universities %
who we expect will be able to create and assess technical topics in a broader range of disciplines.

Second, we will start a new pilot task to assess automatic cross-language report generation.
With the increasing popularity of large generative language models, we expect that
automated report generation over retrieved documents will become increasingly feasible.
We will run a pilot task for generating a multi-paragraph English report from Chinese or Russian documents.
The task will provide a Chinese or Russian document collection
and a set of English report requests describing the English report to be generated.
To broaden the appeal of the task to researchers outside of information retrieval,
participants will 
be given access to a CLIR system for the collection through an API if they do not have their own CLIR system for the selected language.
A major goal of this new pilot task will be to develop an affordable, repeatable and insightful method for scoring the resulting reports.

Finally, we will push back the submission deadline to August for all tasks.
Participation in the 2023 NeuCLIR track was down from participation in the 2022 track.
A poll of teams that submitted in 2022 but not in 2023 revealed that the deadline for submissions was too early.

\subsection{What's Not Changing}

NeuCLIR 2024 will carry over all tasks from NeuCLIR 2023, including ad-hoc cross-language and multilingual retrieval tasks. 

NeuCLIR 2024 will keep the same three-language CLIR and MLIR document collections as in previous years.
To improve the range of tasks for which these collections will be useful,
we also expect to develop and semonstrate a process for producing high-quality topic translations
into additional languages beyond English, Chinese, Persian and Russian.  We also plan to create a repository for storing such translations,
so that future users of the collection can produce CLIR and MLIR runs starting from any existing topic languages, and so that researchers who are interested in additional topic languages can share their topic translations with others.

The NeuCLIR 2022 and 2023 evaluation data,
which includes
sets of one hundred topics with relevance judgments for each of the three languages,
can serve as development sets for year three of the track.
We plan to develop new topic sets in the three NeuCLIR languages
(Chinese, Persian and Russian),
including at least 50 topics per language with relevance judgments, with all topics having judged relevant documents in at least two languages.

%% file: 10-conclusion.tex
\section{Conclusion}

In this second iteration of the TREC NeuCLIR track,
we built upon the success of the first iteration by exploring two additional search tasks:
\textit{multi}-language news search (as a full task), and cross-language search over technical content (as a pilot task).
In total, TREC NeuCLIR 2023 generated 220 runs from six teams.
We observed impressively strong cross-language and multi-language news search effectiveness.
Meanwhile, cross-language information retrieval over technical documents remains challenging,
with no submission achieving an nDCG@20 above 0.5.

These observations motivate the TREC NeuCLIR 2024 tasks.
First, we will continue to benchmark news search in multi-language and cross-language tasks.
For these tasks, we will aim to develop topic sets that include topics that are increasingly challenging for existing systems.
Second, we will elevate our cross-language technical document pilot to a full task
to give this challenging setting more attention
and to construct a more comprehensive benchmark.
And third, we will introduce a new pilot task---cross-language report generation---in which systems will
generate an English report from non-English source material.
There is certainly a lot to look forward to for next year!

%% file: _figs_overlap.tex
\begin{figure*}
    \centering
    \includegraphics[width=0.85\linewidth]{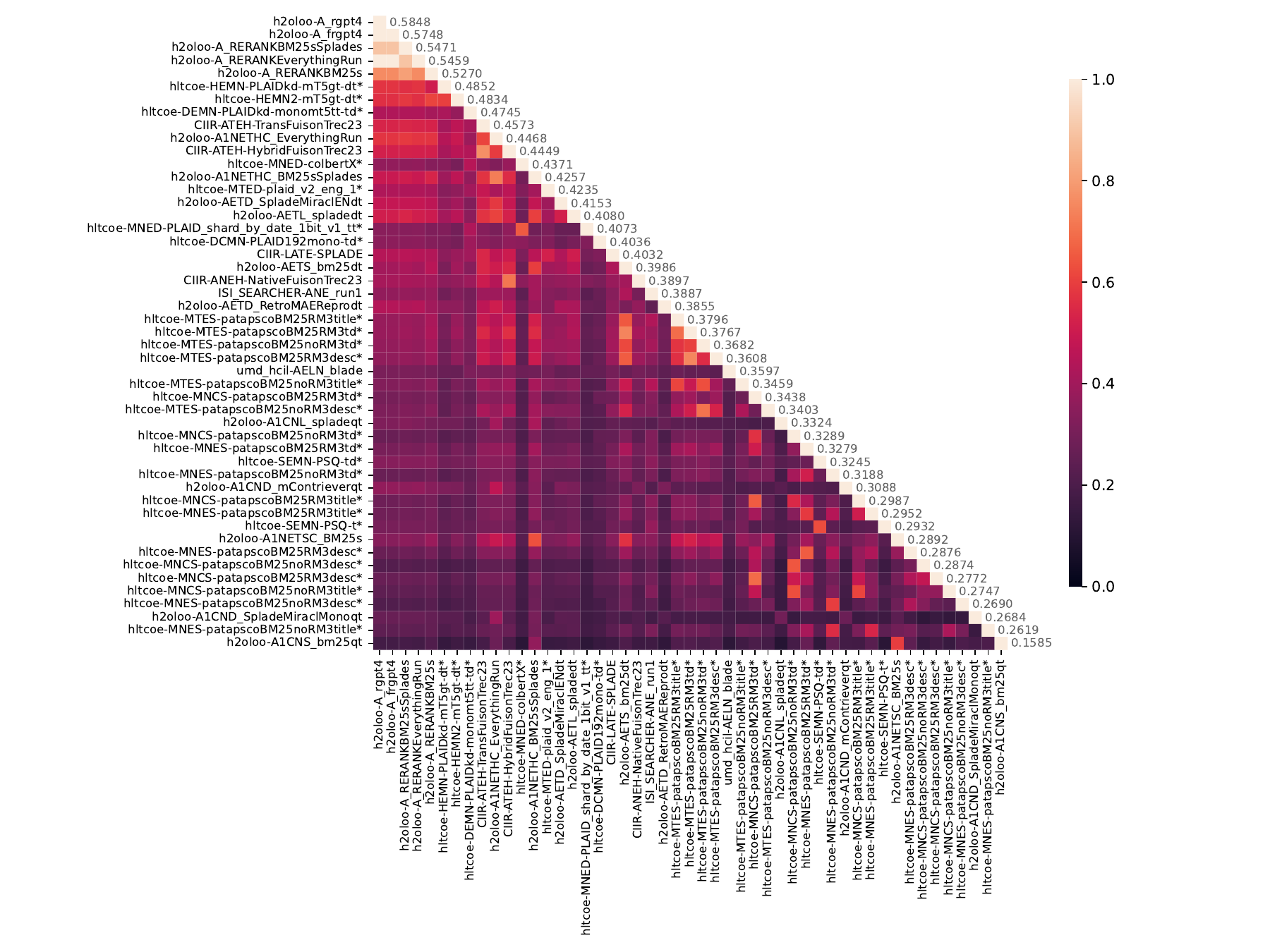}
    \includegraphics[width=0.85\linewidth]{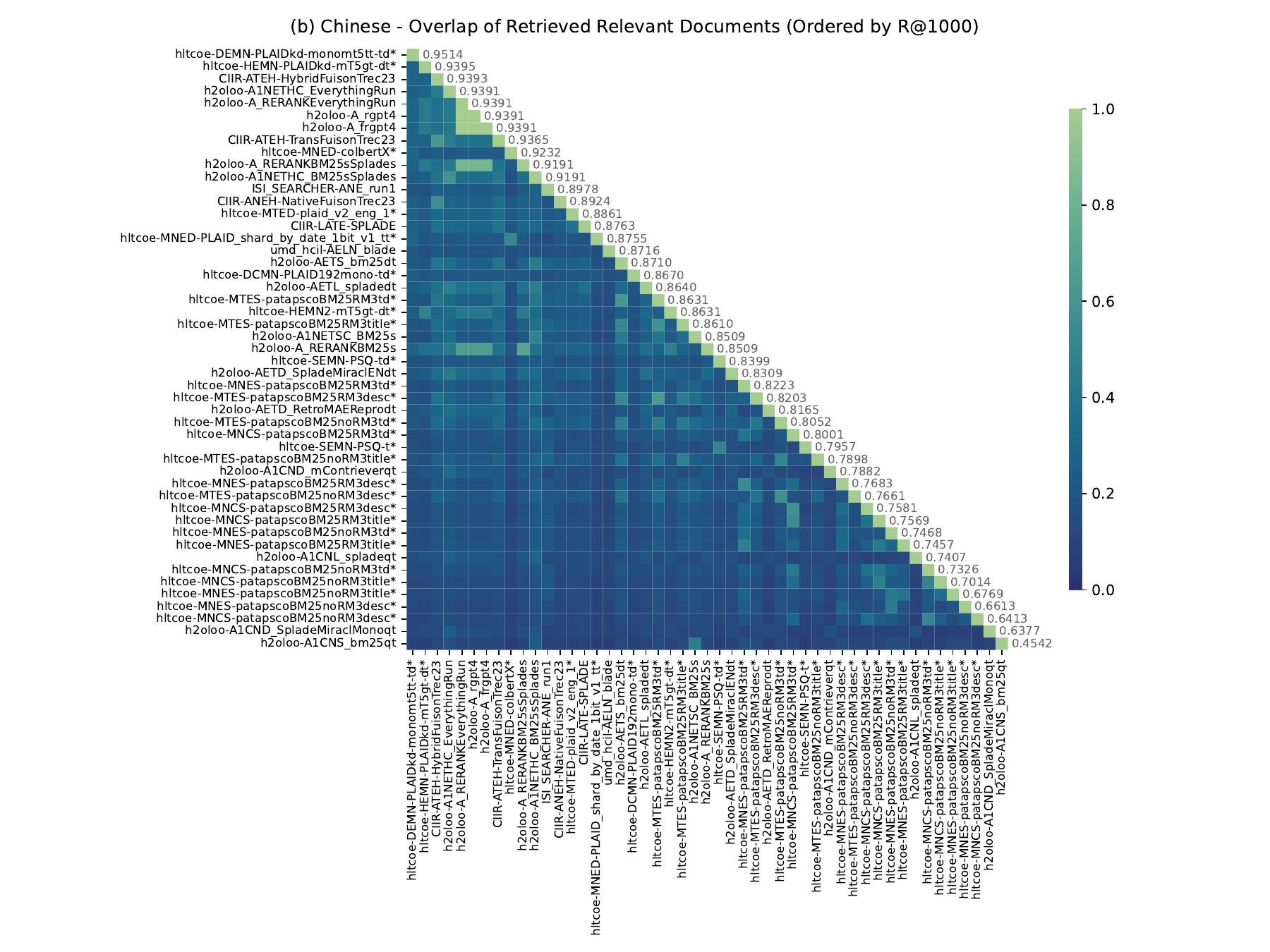}
    \caption{Overlap of documents retrieved by systems that participated in Chinese. * indicates manual runs.}
    \label{fig:zho-overlap}
\end{figure*}

\begin{figure*}
    \centering
    \includegraphics[width=0.85\linewidth]{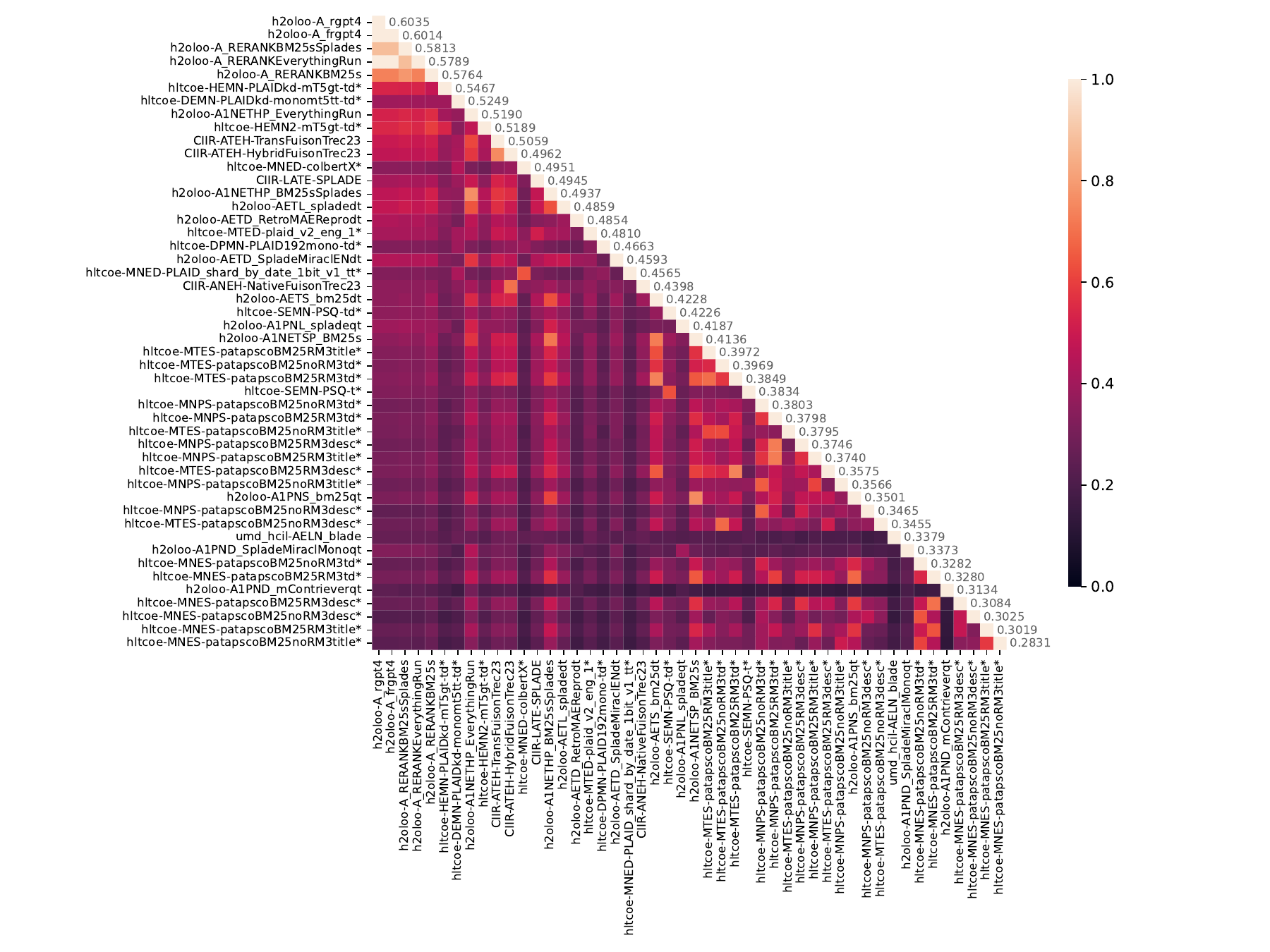}
    \includegraphics[width=0.85\linewidth]{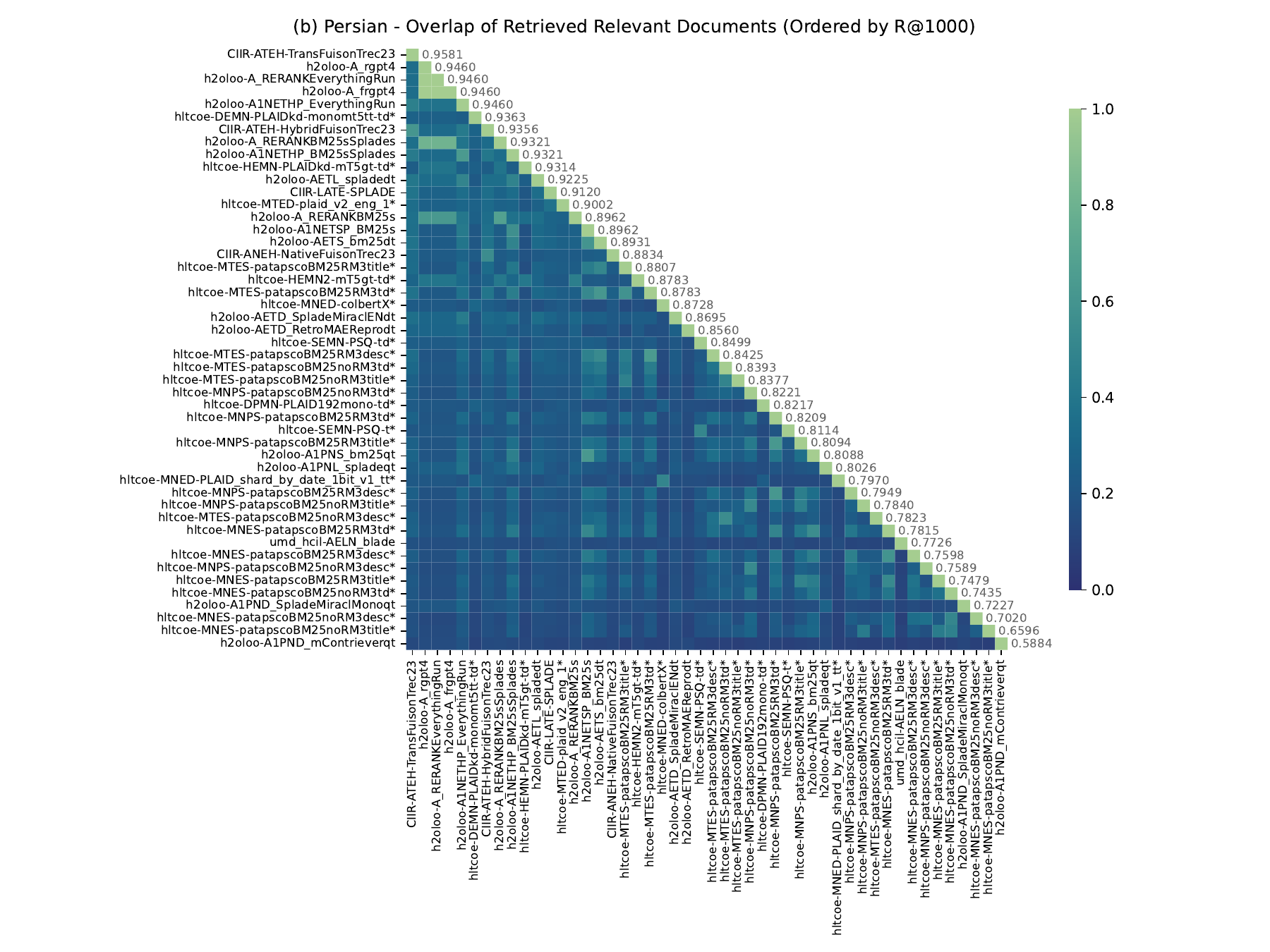}
    \caption{Overlap of documents retrieved by systems that participated in Persian. * indicates manual runs.}
    \label{fig:fas-overlap}
\end{figure*}

\begin{figure*}
    \centering
    \includegraphics[width=0.85\linewidth]{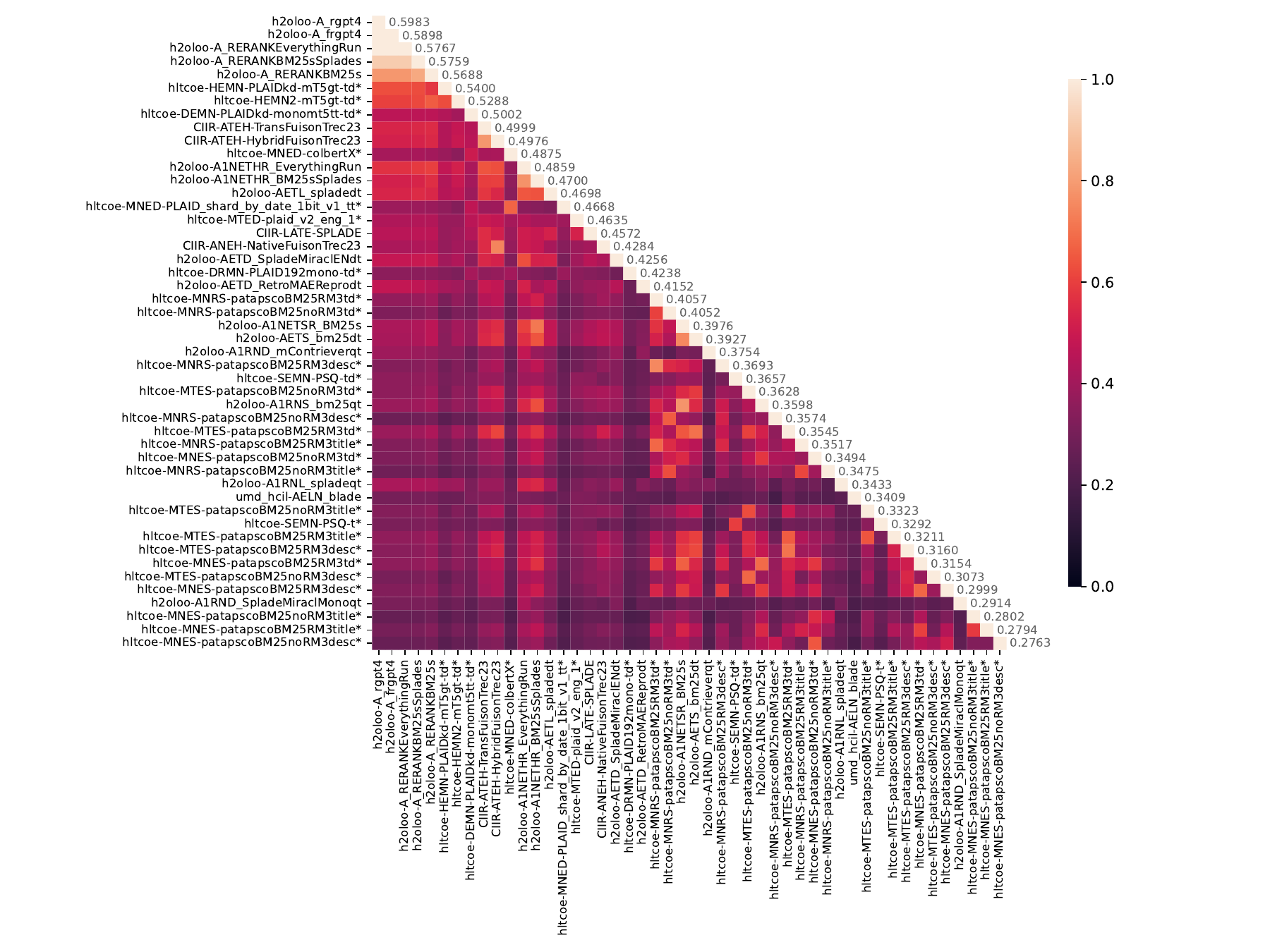}
    \includegraphics[width=0.85\linewidth]{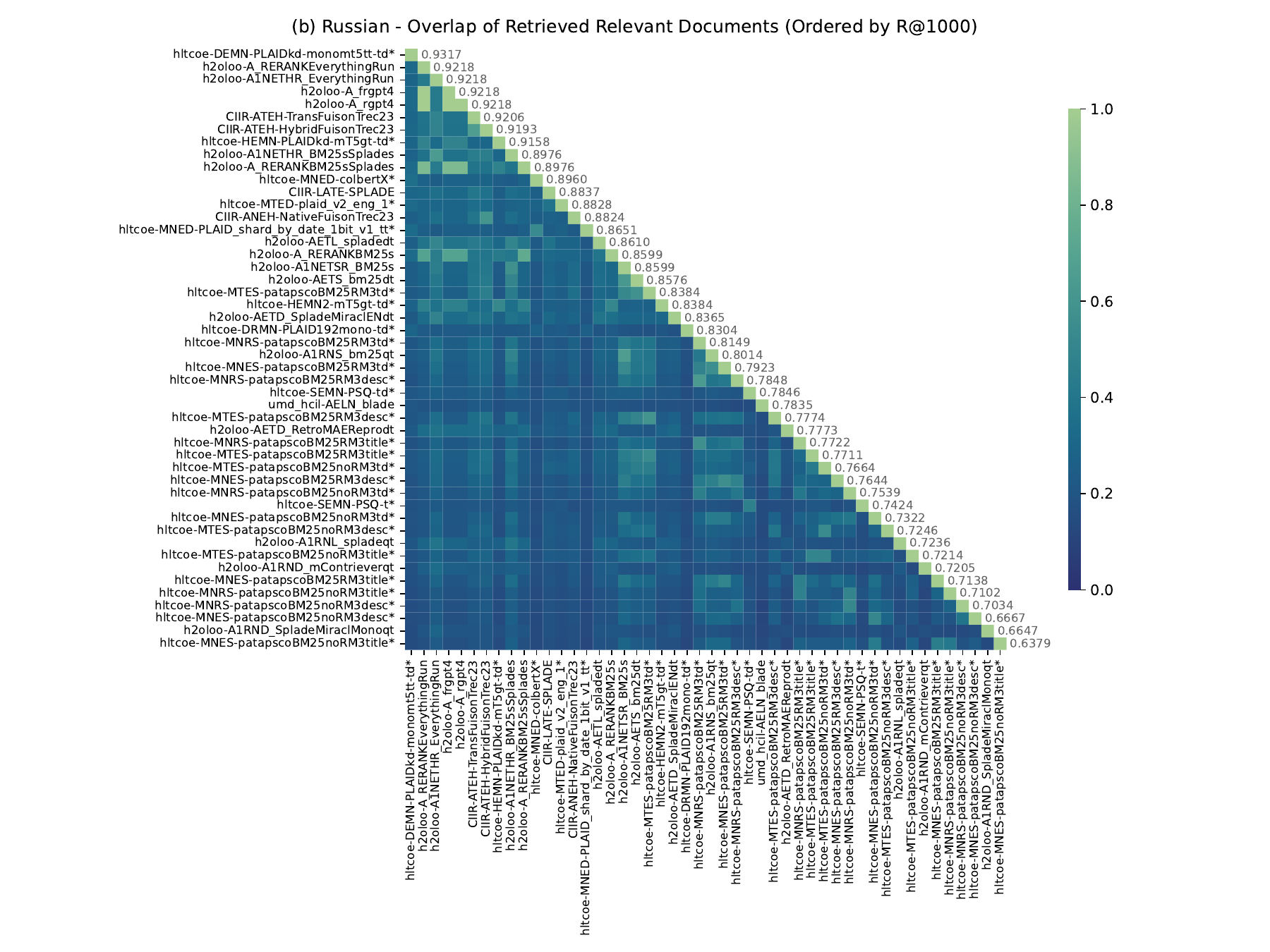}
    \caption{Overlap of documents retrieved by systems that participated in Russian. * indicates manual runs.}
    \label{fig:rus-overlap}
\end{figure*}

\begin{figure*}
    \centering
    \includegraphics[width=0.85\linewidth]{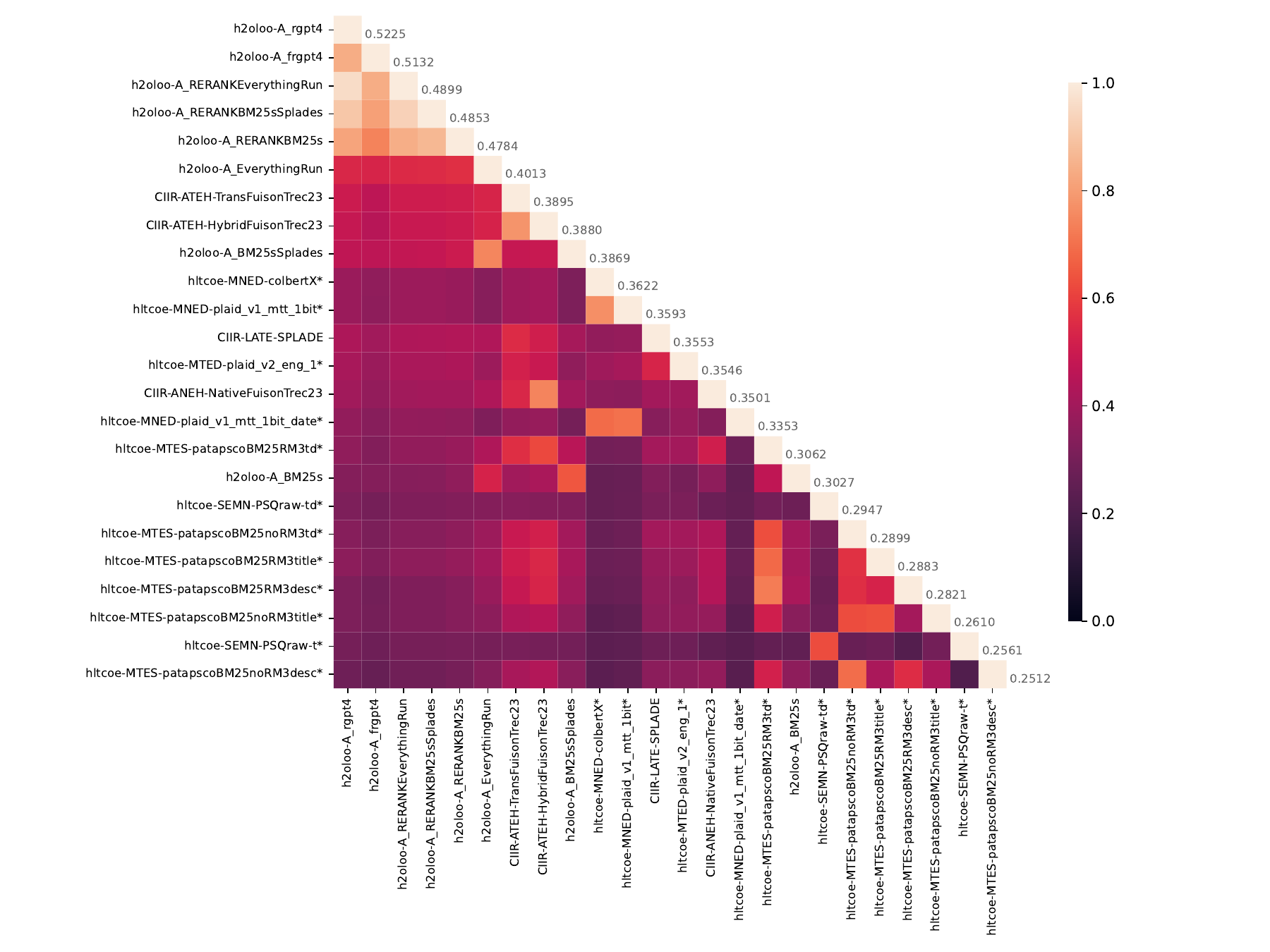}
    \includegraphics[width=0.85\linewidth]{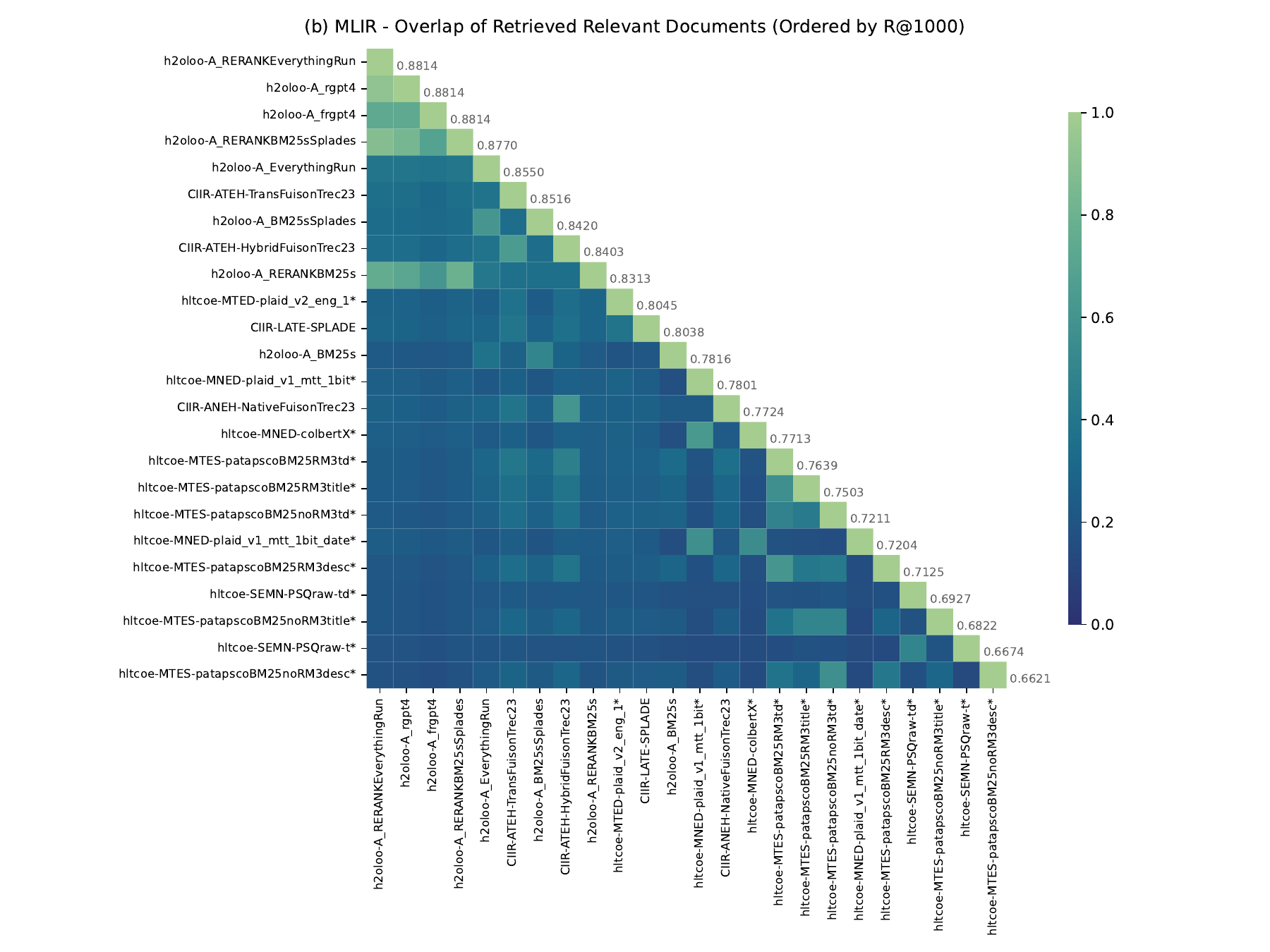}
    \caption{Overlap of documents retrieved by systems that participated in MLIR runs. * indicates manual runs.}
    \label{fig:mlir-overlap}
\end{figure*}

\begin{figure*}
    \centering
    \includegraphics[width=0.85\linewidth]{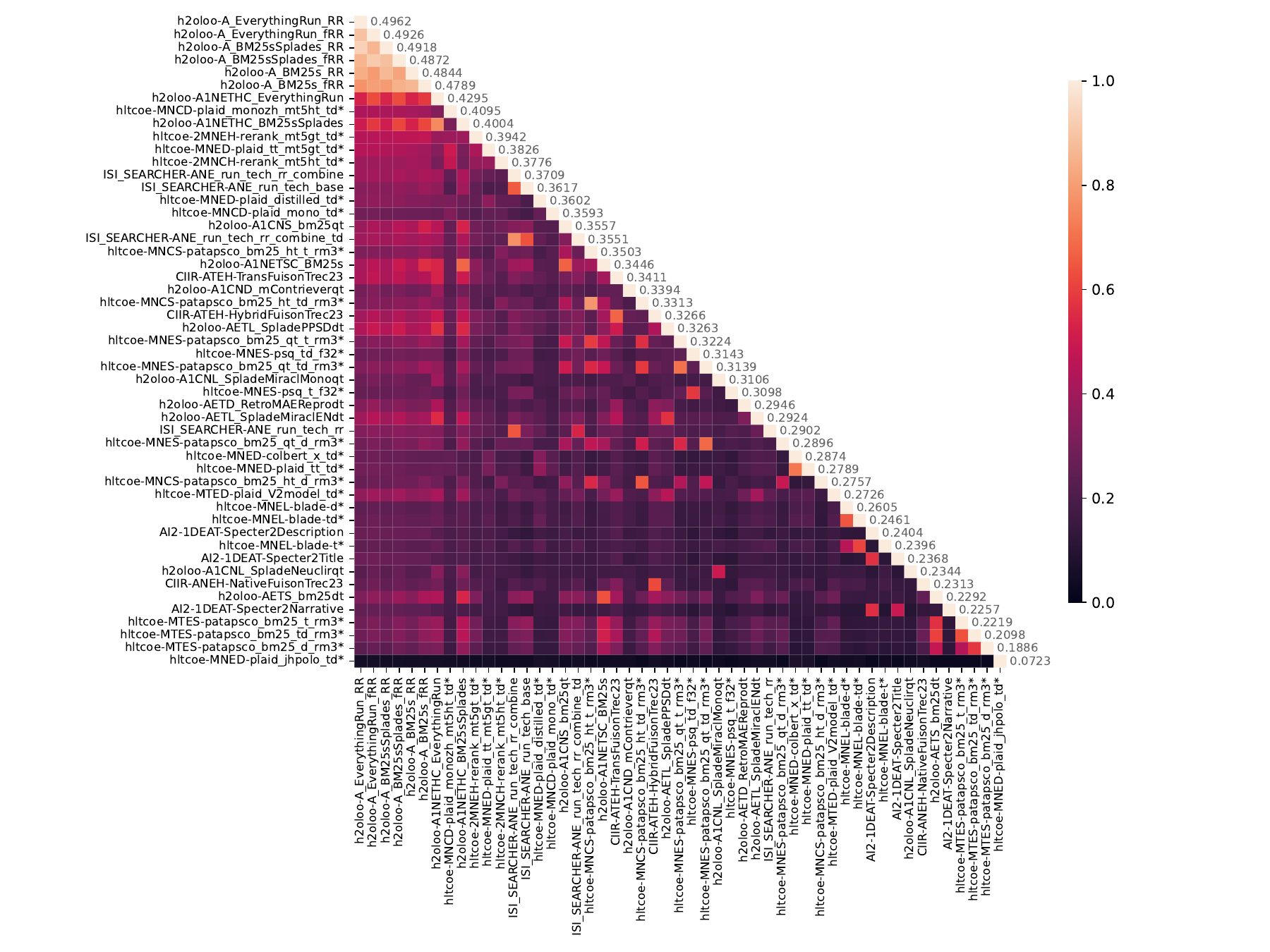}
    \includegraphics[width=0.85\linewidth]{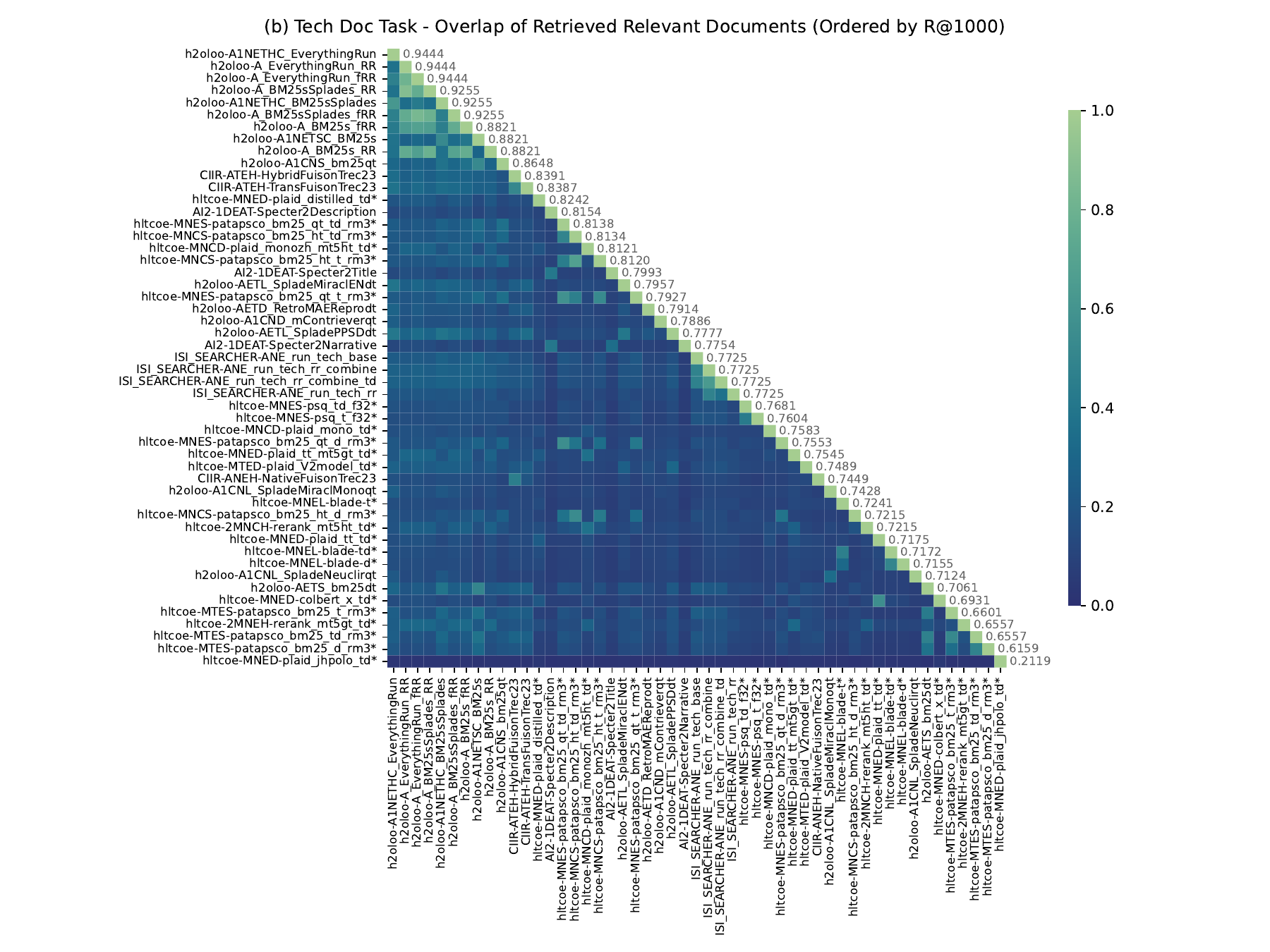}
    \caption{Overlap of documents retrieved by systems that participated in the Technical Document Task. * indicates manual runs.}
    \label{fig:tech-overlap}
\end{figure*}

%% file: _table_clir_full_results.tex
\begin{table*}
\setlength\tabcolsep{0.4em}

\caption{Chinese Results.
Monolingual runs, which use human translations of the queries, are shown in green.
The run used as the first stage retrieval for the reranking task is marked in bold. 
* indicates manual runs. 
Column ``JFD'' indicates whether the run is \underline{j}udged at \underline{f}ull \underline{d}epth, which is 50. 
}\label{tab:zho-full-results}
\centering
\begin{tabular}{ll|ccccc|ccccc}
\toprule
                                         Team &                              Run Name &      ReR. &    Model &       DL &         QL &       JFD  &    nDCG &   RBP &    AP & R@100 &  R@1k \\
\midrule
         naverloo\cite{participants-naverloo} &                                 rgpt4 &    \xmark &   Hybrid &   MT+Org &          E &     \cmark &   0.585 & 0.536 & 0.507 & 0.784 & 0.939 \\
         naverloo\cite{participants-naverloo} &                                frgpt4 &    \xmark &   Hybrid &   MT+Org &          E &     \cmark &   0.575 & 0.518 & 0.496 & 0.784 & 0.939 \\
         naverloo\cite{participants-naverloo} &                    RERANKBM25sSplades &    \xmark &   Hybrid &   MT+Org &          E &     \xmark &   0.547 & 0.486 & 0.465 & 0.784 & 0.919 \\
         naverloo\cite{participants-naverloo} &                   RERANKEverythingRun &    \xmark &   Hybrid &   MT+Org &          E &     \cmark &   0.546 & 0.485 & 0.464 & 0.784 & 0.939 \\
         naverloo\cite{participants-naverloo} &                           RERANKBM25s &    \xmark &   Hybrid &   MT+Org &          E &     \xmark &   0.527 & 0.472 & 0.448 & 0.734 & 0.851 \\
ISI\_SEARCHER\cite{participants-ISI_SEARCHER} &                                  run2 &    \xmark &   Hybrid &      Org &          E &     \xmark &   0.492 & 0.423 & 0.405 & 0.750 & 0.898 \\
             hltcoe\cite{participants-hltcoe} &                     PLAIDkd-mT5gt-dt* &    \xmark &   Hybrid &      Org &          E &     \xmark &   0.485 & 0.432 & 0.406 & 0.730 & 0.940 \\
             hltcoe\cite{participants-hltcoe} &                             mT5gt-dt* &    \cmark &   Hybrid &      Org &          E &     \xmark &   0.483 & 0.430 & 0.399 & 0.709 & 0.863 \\
             hltcoe\cite{participants-hltcoe} &                 PLAIDkd-monomt5tt-td* &    \xmark &    Dense &      Org &          E &     \cmark &   0.474 & 0.411 & 0.385 & 0.749 & 0.951 \\
                 CIIR\cite{participants-CIIR} &                     TransFuisonTrec23 &    \xmark &   Hybrid &       MT &          E &     \cmark &   0.457 & 0.404 & 0.370 & 0.731 & 0.936 \\
         naverloo\cite{participants-naverloo} &                         EverythingRun &    \xmark &   Hybrid &   MT+Org &          E &     \xmark &   0.447 & 0.397 & 0.369 & 0.704 & 0.939 \\
                 CIIR\cite{participants-CIIR} &                    HybridFuisonTrec23 &    \xmark &   Hybrid &       MT &          E &     \cmark &   0.445 & 0.391 & 0.367 & 0.741 & 0.939 \\
             hltcoe\cite{participants-hltcoe} &                             colbertX* &    \xmark &    Dense &      Org &          E &     \cmark &   0.437 & 0.400 & 0.346 & 0.685 & 0.923 \\
         naverloo\cite{participants-naverloo} &                          BM25sSplades &    \xmark &   Hybrid &   MT+Org &          E &     \xmark &   0.426 & 0.381 & 0.358 & 0.671 & 0.919 \\
             hltcoe\cite{participants-hltcoe} &                    plaid\_v2\_eng\_1* &    \xmark &    Dense &       MT &          E &     \cmark &   0.423 & 0.379 & 0.346 & 0.671 & 0.886 \\
         naverloo\cite{participants-naverloo} &                      SpladeMiraclENdt &    \xmark &    Dense &       MT &          E &     \xmark &   0.415 & 0.387 & 0.315 & 0.612 & 0.831 \\
         naverloo\cite{participants-naverloo} &                              spladedt &    \xmark & L-Sparse &       MT &          E &     \xmark &   0.408 & 0.384 & 0.327 & 0.660 & 0.864 \\
             hltcoe\cite{participants-hltcoe} & PLAID\_shard\_by\_date\_1bit\_v1\_tt* &    \xmark &    Dense &      Org &          E &     \cmark &   0.407 & 0.376 & 0.322 & 0.674 & 0.875 \\
             hltcoe\cite{participants-hltcoe} &                 \ml{PLAID192mono-td*} &    \xmark &    Dense &      Org &          C &     \cmark &   0.404 & 0.337 & 0.332 & 0.660 & 0.867 \\
                 CIIR\cite{participants-CIIR} &                                SPLADE &    \xmark & L-Sparse &       MT &          E &     \xmark &   0.403 & 0.354 & 0.330 & 0.653 & 0.876 \\
         naverloo\cite{participants-naverloo} &                                bm25dt &    \xmark &   Sparse &       MT &          E &     \xmark &   0.399 & 0.368 & 0.329 & 0.643 & 0.871 \\
                 CIIR\cite{participants-CIIR} &                    NativeFuisonTrec23 &    \xmark &   Hybrid &      Org &          E &     \cmark &   0.390 & 0.355 & 0.317 & 0.682 & 0.892 \\
ISI\_SEARCHER\cite{participants-ISI_SEARCHER} &                                  run1 &    \xmark & L-Sparse &      Org &          E &     \cmark &   0.389 & 0.351 & 0.326 & 0.685 & 0.898 \\
         naverloo\cite{participants-naverloo} &                       RetroMAEReprodt &    \xmark &    Dense &       MT &          E &     \xmark &   0.386 & 0.342 & 0.289 & 0.616 & 0.817 \\
             hltcoe\cite{participants-hltcoe} &                 patapscoBM25RM3title* &    \xmark &   Sparse &       MT &          E &     \xmark &   0.380 & 0.339 & 0.299 & 0.600 & 0.861 \\
             hltcoe\cite{participants-hltcoe} &               \bf{patapscoBM25RM3td*} &    \xmark &   Sparse &       MT &          E &     \cmark &   0.377 & 0.341 & 0.313 & 0.634 & 0.863 \\
             hltcoe\cite{participants-hltcoe} &                  patapscoBM25noRM3td* &    \xmark &   Sparse &       MT &          E &     \xmark &   0.368 & 0.332 & 0.291 & 0.610 & 0.805 \\
             hltcoe\cite{participants-hltcoe} &                  patapscoBM25RM3desc* &    \xmark &   Sparse &       MT &          E &     \xmark &   0.361 & 0.326 & 0.300 & 0.600 & 0.820 \\
        umd\_hcil\cite{participants-umd_hcil} &                                 blade &    \xmark & L-Sparse &      Org &          E &     \cmark &   0.360 & 0.319 & 0.285 & 0.622 & 0.872 \\
             hltcoe\cite{participants-hltcoe} &               patapscoBM25noRM3title* &    \xmark &   Sparse &       MT &          E &     \xmark &   0.346 & 0.308 & 0.268 & 0.574 & 0.790 \\
             hltcoe\cite{participants-hltcoe} &               \ml{patapscoBM25RM3td*} &    \xmark &   Sparse &      Org &          C &     \xmark &   0.344 & 0.313 & 0.267 & 0.551 & 0.800 \\
             hltcoe\cite{participants-hltcoe} &                patapscoBM25noRM3desc* &    \xmark &   Sparse &       MT &          E &     \xmark &   0.340 & 0.306 & 0.275 & 0.588 & 0.766 \\
         naverloo\cite{participants-naverloo} &                              spladeqt &    \xmark & L-Sparse &      Org &          E &     \xmark &   0.332 & 0.289 & 0.251 & 0.502 & 0.741 \\
             hltcoe\cite{participants-hltcoe} &             \ml{patapscoBM25noRM3td*} &    \xmark &   Sparse &      Org &          C &     \xmark &   0.329 & 0.302 & 0.240 & 0.503 & 0.733 \\
             hltcoe\cite{participants-hltcoe} &                    patapscoBM25RM3td* &    \xmark &   Sparse &      Org &          E &     \xmark &   0.328 & 0.307 & 0.254 & 0.563 & 0.822 \\
             hltcoe\cite{participants-hltcoe} &                               PSQ-td* &    \xmark &   Sparse &      Org &          E &     \cmark &   0.324 & 0.294 & 0.264 & 0.624 & 0.840 \\
             hltcoe\cite{participants-hltcoe} &                  patapscoBM25noRM3td* &    \xmark &   Sparse &      Org &          E &     \xmark &   0.319 & 0.291 & 0.227 & 0.491 & 0.747 \\
         naverloo\cite{participants-naverloo} &                         mContrieverqt &    \xmark &    Dense &      Org &          E &     \xmark &   0.309 & 0.282 & 0.219 & 0.552 & 0.788 \\
             hltcoe\cite{participants-hltcoe} &            \ml{patapscoBM25RM3title*} &    \xmark &   Sparse &      Org &          C &     \xmark &   0.299 & 0.265 & 0.234 & 0.513 & 0.757 \\
             hltcoe\cite{participants-hltcoe} &                 patapscoBM25RM3title* &    \xmark &   Sparse &      Org &          E &     \xmark &   0.295 & 0.256 & 0.226 & 0.508 & 0.746 \\
             hltcoe\cite{participants-hltcoe} &                                PSQ-t* &    \xmark &   Sparse &      Org &          E &     \xmark &   0.293 & 0.255 & 0.226 & 0.554 & 0.796 \\
         naverloo\cite{participants-naverloo} &                                 BM25s &    \xmark &   Sparse &   MT+Org &          E &     \xmark &   0.289 & 0.269 & 0.230 & 0.576 & 0.851 \\
             hltcoe\cite{participants-hltcoe} &                  patapscoBM25RM3desc* &    \xmark &   Sparse &      Org &          E &     \xmark &   0.288 & 0.261 & 0.225 & 0.516 & 0.768 \\
             hltcoe\cite{participants-hltcoe} &           \ml{patapscoBM25noRM3desc*} &    \xmark &   Sparse &      Org &          C &     \xmark &   0.287 & 0.257 & 0.208 & 0.441 & 0.641 \\
             hltcoe\cite{participants-hltcoe} &             \ml{patapscoBM25RM3desc*} &    \xmark &   Sparse &      Org &          C &     \xmark &   0.277 & 0.257 & 0.212 & 0.467 & 0.758 \\
             hltcoe\cite{participants-hltcoe} &          \ml{patapscoBM25noRM3title*} &    \xmark &   Sparse &      Org &          C &     \xmark &   0.275 & 0.253 & 0.201 & 0.463 & 0.701 \\
             hltcoe\cite{participants-hltcoe} &                patapscoBM25noRM3desc* &    \xmark &   Sparse &      Org &          E &     \xmark &   0.269 & 0.233 & 0.185 & 0.412 & 0.661 \\
         naverloo\cite{participants-naverloo} &                    SpladeMiraclMonoqt &    \xmark &    Dense &      Org &          E &     \xmark &   0.268 & 0.237 & 0.194 & 0.441 & 0.638 \\
             hltcoe\cite{participants-hltcoe} &               patapscoBM25noRM3title* &    \xmark &   Sparse &      Org &          E &     \xmark &   0.262 & 0.229 & 0.183 & 0.444 & 0.677 \\
         naverloo\cite{participants-naverloo} &                                bm25qt &    \xmark &   Sparse &      Org &          E &     \xmark &   0.159 & 0.164 & 0.100 & 0.263 & 0.454 \\
\bottomrule
\end{tabular}

\begin{flushleft}
\footnotesize{$\dagger$ \texttt{run2} from \texttt{ISI\_SEARCHER} is a late submission to NIST and was not included in the pool.} \\
\footnotesize{Team \texttt{h2oloo} is renamed as \texttt{naverloo} to reflect the participating parties.}
\end{flushleft}
\end{table*}

\begin{table*}[]
\setlength\tabcolsep{0.5em}

\caption{Persian Results.
Monolingual runs, which use human translations of the queries, are marked as green.
Run used as the first stage retrieval for the reranking task is marked bold. 
* indicates manual runs.
Column ``JFD'' indicates whether the run is \underline{j}udged with \underline{f}ull \underline{d}epth, which is 50. 
}\label{tab:fas-full-results}
    \centering

\begin{tabular}{ll|ccccc|ccccc}
\toprule
                                 Team &                              Run Name &      ReR. &    Model &       DL &         QL &       JFD  &    nDCG &   RBP &    AP & R@100 &  R@1k \\
\midrule
 naverloo\cite{participants-naverloo} &                                 rgpt4 &    \xmark &   Hybrid &   MT+Org &          E &     \cmark &   0.603 & 0.520 & 0.521 & 0.761 & 0.946 \\
 naverloo\cite{participants-naverloo} &                                frgpt4 &    \xmark &   Hybrid &   MT+Org &          E &     \cmark &   0.601 & 0.511 & 0.519 & 0.761 & 0.946 \\
 naverloo\cite{participants-naverloo} &                    RERANKBM25sSplades &    \xmark &   Hybrid &   MT+Org &          E &     \xmark &   0.581 & 0.484 & 0.502 & 0.772 & 0.932 \\
 naverloo\cite{participants-naverloo} &                   RERANKEverythingRun &    \xmark &   Hybrid &   MT+Org &          E &     \cmark &   0.579 & 0.482 & 0.500 & 0.761 & 0.946 \\
 naverloo\cite{participants-naverloo} &                           RERANKBM25s &    \xmark &   Hybrid &   MT+Org &          E &     \xmark &   0.576 & 0.479 & 0.495 & 0.769 & 0.896 \\
     hltcoe\cite{participants-hltcoe} &                     PLAIDkd-mT5gt-td* &    \xmark &   Hybrid &      Org &          E &     \xmark &   0.547 & 0.468 & 0.450 & 0.743 & 0.931 \\
     hltcoe\cite{participants-hltcoe} &                 PLAIDkd-monomt5tt-td* &    \xmark &    Dense &      Org &          E &     \cmark &   0.525 & 0.453 & 0.461 & 0.758 & 0.936 \\
     hltcoe\cite{participants-hltcoe} &                             mT5gt-td* &    \cmark &   Hybrid &      Org &          E &     \cmark &   0.519 & 0.443 & 0.440 & 0.722 & 0.878 \\
 naverloo\cite{participants-naverloo} &                         EverythingRun &    \xmark &   Hybrid &   MT+Org &          E &     \xmark &   0.519 & 0.423 & 0.447 & 0.795 & 0.946 \\
         CIIR\cite{participants-CIIR} &                     TransFuisonTrec23 &    \xmark &   Hybrid &       MT &          E &     \cmark &   0.506 & 0.413 & 0.446 & 0.752 & 0.958 \\
         CIIR\cite{participants-CIIR} &                    HybridFuisonTrec23 &    \xmark &   Hybrid &       MT &          E &     \cmark &   0.496 & 0.414 & 0.438 & 0.733 & 0.936 \\
     hltcoe\cite{participants-hltcoe} &                             colbertX* &    \xmark &    Dense &      Org &          E &     \cmark &   0.495 & 0.432 & 0.421 & 0.666 & 0.873 \\
         CIIR\cite{participants-CIIR} &                                SPLADE &    \xmark & L-Sparse &       MT &          E &     \xmark &   0.495 & 0.411 & 0.411 & 0.707 & 0.912 \\
 naverloo\cite{participants-naverloo} &                          BM25sSplades &    \xmark &   Hybrid &   MT+Org &          E &     \xmark &   0.494 & 0.410 & 0.433 & 0.779 & 0.932 \\
 naverloo\cite{participants-naverloo} &                              spladedt &    \xmark & L-Sparse &       MT &          E &     \xmark &   0.486 & 0.422 & 0.407 & 0.739 & 0.922 \\
 naverloo\cite{participants-naverloo} &                       RetroMAEReprodt &    \xmark &    Dense &       MT &          E &     \xmark &   0.485 & 0.394 & 0.379 & 0.645 & 0.856 \\
     hltcoe\cite{participants-hltcoe} &                    plaid\_v2\_eng\_1* &    \xmark &    Dense &       MT &          E &     \cmark &   0.481 & 0.417 & 0.421 & 0.693 & 0.900 \\
     hltcoe\cite{participants-hltcoe} &                 \ml{PLAID192mono-td*} &    \xmark &    Dense &      Org &          P &     \cmark &   0.466 & 0.405 & 0.399 & 0.620 & 0.822 \\
 naverloo\cite{participants-naverloo} &                      SpladeMiraclENdt &    \xmark &    Dense &       MT &          E &     \xmark &   0.459 & 0.394 & 0.368 & 0.667 & 0.870 \\
     hltcoe\cite{participants-hltcoe} & PLAID\_shard\_by\_date\_1bit\_v1\_tt* &    \xmark &    Dense &      Org &          E &     \cmark &   0.456 & 0.411 & 0.392 & 0.595 & 0.797 \\
         CIIR\cite{participants-CIIR} &                    NativeFuisonTrec23 &    \xmark &   Hybrid &      Org &          E &     \cmark &   0.440 & 0.364 & 0.355 & 0.659 & 0.883 \\
 naverloo\cite{participants-naverloo} &                                bm25dt &    \xmark &   Sparse &       MT &          E &     \xmark &   0.423 & 0.348 & 0.373 & 0.702 & 0.893 \\
     hltcoe\cite{participants-hltcoe} &                               PSQ-td* &    \xmark &   Sparse &      Org &          E &     \cmark &   0.423 & 0.340 & 0.348 & 0.628 & 0.850 \\
 naverloo\cite{participants-naverloo} &                              spladeqt &    \xmark & L-Sparse &      Org &          E &     \xmark &   0.419 & 0.351 & 0.333 & 0.608 & 0.803 \\
 naverloo\cite{participants-naverloo} &                                 BM25s &    \xmark &   Sparse &   MT+Org &          E &     \xmark &   0.414 & 0.349 & 0.370 & 0.696 & 0.896 \\
     hltcoe\cite{participants-hltcoe} &                 patapscoBM25RM3title* &    \xmark &   Sparse &       MT &          E &     \xmark &   0.397 & 0.344 & 0.341 & 0.688 & 0.881 \\
     hltcoe\cite{participants-hltcoe} &                  patapscoBM25noRM3td* &    \xmark &   Sparse &       MT &          E &     \xmark &   0.397 & 0.322 & 0.337 & 0.654 & 0.839 \\
     hltcoe\cite{participants-hltcoe} &               \bf{patapscoBM25RM3td*} &    \xmark &   Sparse &       MT &          E &     \cmark &   0.385 & 0.333 & 0.340 & 0.654 & 0.878 \\
     hltcoe\cite{participants-hltcoe} &                                PSQ-t* &    \xmark &   Sparse &      Org &          E &     \xmark &   0.383 & 0.330 & 0.313 & 0.597 & 0.811 \\
     hltcoe\cite{participants-hltcoe} &             \ml{patapscoBM25noRM3td*} &    \xmark &   Sparse &      Org &          P &     \xmark &   0.380 & 0.308 & 0.313 & 0.595 & 0.822 \\
     hltcoe\cite{participants-hltcoe} &               \ml{patapscoBM25RM3td*} &    \xmark &   Sparse &      Org &          P &     \xmark &   0.380 & 0.320 & 0.330 & 0.622 & 0.821 \\
     hltcoe\cite{participants-hltcoe} &               patapscoBM25noRM3title* &    \xmark &   Sparse &       MT &          E &     \xmark &   0.379 & 0.334 & 0.303 & 0.606 & 0.838 \\
     hltcoe\cite{participants-hltcoe} &             \ml{patapscoBM25RM3desc*} &    \xmark &   Sparse &      Org &          P &     \xmark &   0.375 & 0.315 & 0.324 & 0.567 & 0.795 \\
     hltcoe\cite{participants-hltcoe} &            \ml{patapscoBM25RM3title*} &    \xmark &   Sparse &      Org &          P &     \xmark &   0.374 & 0.312 & 0.321 & 0.612 & 0.809 \\
     hltcoe\cite{participants-hltcoe} &          \ml{patapscoBM25noRM3title*} &    \xmark &   Sparse &      Org &          P &     \xmark &   0.357 & 0.302 & 0.280 & 0.566 & 0.784 \\
     hltcoe\cite{participants-hltcoe} &                  patapscoBM25RM3desc* &    \xmark &   Sparse &       MT &          E &     \xmark &   0.357 & 0.294 & 0.308 & 0.605 & 0.842 \\
 naverloo\cite{participants-naverloo} &                                bm25qt &    \xmark &   Sparse &      Org &          E &     \xmark &   0.350 & 0.292 & 0.301 & 0.598 & 0.809 \\
     hltcoe\cite{participants-hltcoe} &           \ml{patapscoBM25noRM3desc*} &    \xmark &   Sparse &      Org &          P &     \xmark &   0.347 & 0.275 & 0.291 & 0.568 & 0.759 \\
     hltcoe\cite{participants-hltcoe} &                patapscoBM25noRM3desc* &    \xmark &   Sparse &       MT &          E &     \xmark &   0.346 & 0.270 & 0.296 & 0.587 & 0.782 \\
umd\_hcil\cite{participants-umd_hcil} &                                 blade &    \xmark & L-Sparse &      Org &          E &     \cmark &   0.338 & 0.287 & 0.264 & 0.548 & 0.773 \\
 naverloo\cite{participants-naverloo} &                    SpladeMiraclMonoqt &    \xmark &    Dense &      Org &          E &     \xmark &   0.337 & 0.271 & 0.263 & 0.528 & 0.723 \\
     hltcoe\cite{participants-hltcoe} &                  patapscoBM25noRM3td* &    \xmark &   Sparse &      Org &          E &     \xmark &   0.328 & 0.254 & 0.262 & 0.526 & 0.743 \\
     hltcoe\cite{participants-hltcoe} &                    patapscoBM25RM3td* &    \xmark &   Sparse &      Org &          E &     \xmark &   0.328 & 0.267 & 0.282 & 0.565 & 0.781 \\
 naverloo\cite{participants-naverloo} &                         mContrieverqt &    \xmark &    Dense &      Org &          E &     \xmark &   0.313 & 0.234 & 0.213 & 0.415 & 0.588 \\
     hltcoe\cite{participants-hltcoe} &                  patapscoBM25RM3desc* &    \xmark &   Sparse &      Org &          E &     \xmark &   0.308 & 0.257 & 0.265 & 0.527 & 0.760 \\
     hltcoe\cite{participants-hltcoe} &                patapscoBM25noRM3desc* &    \xmark &   Sparse &      Org &          E &     \xmark &   0.303 & 0.237 & 0.244 & 0.500 & 0.702 \\
     hltcoe\cite{participants-hltcoe} &                 patapscoBM25RM3title* &    \xmark &   Sparse &      Org &          E &     \xmark &   0.302 & 0.257 & 0.245 & 0.547 & 0.748 \\
     hltcoe\cite{participants-hltcoe} &               patapscoBM25noRM3title* &    \xmark &   Sparse &      Org &          E &     \xmark &   0.283 & 0.251 & 0.202 & 0.454 & 0.660 \\
\bottomrule
\end{tabular}
\begin{flushleft}
\footnotesize{Team \texttt{h2oloo} is renamed as \texttt{naverloo} to reflect the participating parties.}
\end{flushleft}

\end{table*}

\begin{table*}[]
\setlength\tabcolsep{0.5em}

\caption{Russian Results.
Monolingual runs, which use human translations of the queries, are marked as green.
Run used as the first stage retrieval for the reranking task is marked bold. 
* indicates manual runs.
Column ``JFD'' indicates whether the run is \underline{j}udged with \underline{f}ull \underline{d}epth, which is 50. 
}\label{tab:rus-full-results}
    \centering
\begin{tabular}{ll|ccccc|ccccc}
\toprule
                                  Team &                              Run Name &      ReR. &    Model &       DL &         QL &       JFD  &    nDCG &   RBP &    AP & R@100 &  R@1k \\
\midrule
  naverloo\cite{participants-naverloo} &                                 rgpt4 &    \xmark &   Hybrid &   MT+Org &          E &     \cmark &   0.598 & 0.584 & 0.519 & 0.763 & 0.922 \\
  naverloo\cite{participants-naverloo} &                                frgpt4 &    \xmark &   Hybrid &   MT+Org &          E &     \cmark &   0.590 & 0.569 & 0.509 & 0.763 & 0.922 \\
  naverloo\cite{participants-naverloo} &                   RERANKEverythingRun &    \xmark &   Hybrid &   MT+Org &          E &     \cmark &   0.577 & 0.545 & 0.500 & 0.763 & 0.922 \\
  naverloo\cite{participants-naverloo} &                    RERANKBM25sSplades &    \xmark &   Hybrid &   MT+Org &          E &     \cmark &   0.576 & 0.543 & 0.496 & 0.760 & 0.898 \\
  naverloo\cite{participants-naverloo} &                           RERANKBM25s &    \xmark &   Hybrid &   MT+Org &          E &     \xmark &   0.569 & 0.537 & 0.482 & 0.746 & 0.860 \\
      hltcoe\cite{participants-hltcoe} &                     PLAIDkd-mT5gt-td* &    \xmark &   Hybrid &      Org &          E &     \cmark &   0.540 & 0.501 & 0.466 & 0.738 & 0.916 \\
      hltcoe\cite{participants-hltcoe} &                             mT5gt-td* &    \cmark &   Hybrid &      Org &          E &     \cmark &   0.529 & 0.501 & 0.441 & 0.693 & 0.838 \\
      hltcoe\cite{participants-hltcoe} &                 PLAIDkd-monomt5tt-td* &    \xmark &    Dense &      Org &          E &     \cmark &   0.500 & 0.473 & 0.429 & 0.740 & 0.932 \\
          CIIR\cite{participants-CIIR} &                     TransFuisonTrec23 &    \xmark &   Hybrid &       MT &          E &     \cmark &   0.500 & 0.465 & 0.412 & 0.685 & 0.921 \\
          CIIR\cite{participants-CIIR} &                    HybridFuisonTrec23 &    \xmark &   Hybrid &       MT &          E &     \cmark &   0.498 & 0.458 & 0.408 & 0.679 & 0.919 \\
      hltcoe\cite{participants-hltcoe} &                             colbertX* &    \xmark &    Dense &      Org &          E &     \cmark &   0.488 & 0.454 & 0.399 & 0.696 & 0.896 \\
  naverloo\cite{participants-naverloo} &                         EverythingRun &    \xmark &   Hybrid &   MT+Org &          E &     \xmark &   0.486 & 0.472 & 0.398 & 0.670 & 0.922 \\
  naverloo\cite{participants-naverloo} &                          BM25sSplades &    \xmark &   Hybrid &   MT+Org &          E &     \xmark &   0.470 & 0.455 & 0.387 & 0.671 & 0.898 \\
  naverloo\cite{participants-naverloo} &                              spladedt &    \xmark & L-Sparse &       MT &          E &     \xmark &   0.470 & 0.464 & 0.375 & 0.642 & 0.861 \\
      hltcoe\cite{participants-hltcoe} & PLAID\_shard\_by\_date\_1bit\_v1\_tt* &    \xmark &    Dense &      Org &          E &     \cmark &   0.467 & 0.443 & 0.383 & 0.657 & 0.865 \\
      hltcoe\cite{participants-hltcoe} &                    plaid\_v2\_eng\_1* &    \xmark &    Dense &       MT &          E &     \cmark &   0.463 & 0.425 & 0.386 & 0.678 & 0.883 \\
          CIIR\cite{participants-CIIR} &                                SPLADE &    \xmark & L-Sparse &       MT &          E &     \xmark &   0.457 & 0.419 & 0.369 & 0.654 & 0.884 \\
          CIIR\cite{participants-CIIR} &                    NativeFuisonTrec23 &    \xmark &   Hybrid &      Org &          E &     \cmark &   0.428 & 0.401 & 0.325 & 0.610 & 0.882 \\
  naverloo\cite{participants-naverloo} &                      SpladeMiraclENdt &    \xmark &    Dense &       MT &          E &     \xmark &   0.426 & 0.417 & 0.325 & 0.627 & 0.836 \\
      hltcoe\cite{participants-hltcoe} &                 \ml{PLAID192mono-td*} &    \xmark &    Dense &      Org &          R &     \cmark &   0.424 & 0.400 & 0.347 & 0.596 & 0.830 \\
  naverloo\cite{participants-naverloo} &                       RetroMAEReprodt &    \xmark &    Dense &       MT &          E &     \xmark &   0.415 & 0.411 & 0.322 & 0.568 & 0.777 \\
      hltcoe\cite{participants-hltcoe} &               \ml{patapscoBM25RM3td*} &    \xmark &   Sparse &      Org &          R &     \xmark &   0.406 & 0.364 & 0.306 & 0.580 & 0.815 \\
      hltcoe\cite{participants-hltcoe} &             \ml{patapscoBM25noRM3td*} &    \xmark &   Sparse &      Org &          R &     \xmark &   0.405 & 0.369 & 0.289 & 0.548 & 0.754 \\
  naverloo\cite{participants-naverloo} &                                 BM25s &    \xmark &   Sparse &   MT+Org &          E &     \xmark &   0.398 & 0.396 & 0.323 & 0.622 & 0.860 \\
  naverloo\cite{participants-naverloo} &                                bm25dt &    \xmark &   Sparse &       MT &          E &     \xmark &   0.393 & 0.391 & 0.303 & 0.620 & 0.858 \\
  naverloo\cite{participants-naverloo} &                         mContrieverqt &    \xmark &    Dense &      Org &          E &     \xmark &   0.375 & 0.366 & 0.264 & 0.484 & 0.720 \\
      hltcoe\cite{participants-hltcoe} &             \ml{patapscoBM25RM3desc*} &    \xmark &   Sparse &      Org &          R &     \xmark &   0.369 & 0.339 & 0.276 & 0.549 & 0.785 \\
      hltcoe\cite{participants-hltcoe} &                               PSQ-td* &    \xmark &   Sparse &      Org &          E &     \cmark &   0.366 & 0.351 & 0.286 & 0.554 & 0.785 \\
      hltcoe\cite{participants-hltcoe} &                  patapscoBM25noRM3td* &    \xmark &   Sparse &       MT &          E &     \xmark &   0.363 & 0.357 & 0.283 & 0.573 & 0.766 \\
  naverloo\cite{participants-naverloo} &                                bm25qt &    \xmark &   Sparse &      Org &          E &     \xmark &   0.360 & 0.352 & 0.280 & 0.557 & 0.801 \\
      hltcoe\cite{participants-hltcoe} &           \ml{patapscoBM25noRM3desc*} &    \xmark &   Sparse &      Org &          R &     \xmark &   0.357 & 0.328 & 0.253 & 0.495 & 0.703 \\
      hltcoe\cite{participants-hltcoe} &               \bf{patapscoBM25RM3td*} &    \xmark &   Sparse &       MT &          E &     \cmark &   0.355 & 0.364 & 0.281 & 0.582 & 0.838 \\
      hltcoe\cite{participants-hltcoe} &            \ml{patapscoBM25RM3title*} &    \xmark &   Sparse &      Org &          R &     \xmark &   0.352 & 0.323 & 0.261 & 0.512 & 0.772 \\
      hltcoe\cite{participants-hltcoe} &                  patapscoBM25noRM3td* &    \xmark &   Sparse &      Org &          E &     \xmark &   0.349 & 0.336 & 0.255 & 0.520 & 0.732 \\
      hltcoe\cite{participants-hltcoe} &          \ml{patapscoBM25noRM3title*} &    \xmark &   Sparse &      Org &          R &     \xmark &   0.347 & 0.325 & 0.244 & 0.493 & 0.710 \\
  naverloo\cite{participants-naverloo} &                              spladeqt &    \xmark & L-Sparse &      Org &          E &     \xmark &   0.343 & 0.349 & 0.271 & 0.519 & 0.724 \\
 umd\_hcil\cite{participants-umd_hcil} &                                 blade &    \xmark & L-Sparse &      Org &          E &     \cmark &   0.341 & 0.324 & 0.265 & 0.533 & 0.783 \\
      hltcoe\cite{participants-hltcoe} &               patapscoBM25noRM3title* &    \xmark &   Sparse &       MT &          E &     \xmark &   0.332 & 0.333 & 0.256 & 0.530 & 0.721 \\
      hltcoe\cite{participants-hltcoe} &                                PSQ-t* &    \xmark &   Sparse &      Org &          E &     \xmark &   0.329 & 0.310 & 0.263 & 0.504 & 0.742 \\
      hltcoe\cite{participants-hltcoe} &                 patapscoBM25RM3title* &    \xmark &   Sparse &       MT &          E &     \xmark &   0.321 & 0.320 & 0.257 & 0.534 & 0.771 \\
      hltcoe\cite{participants-hltcoe} &                  patapscoBM25RM3desc* &    \xmark &   Sparse &       MT &          E &     \xmark &   0.316 & 0.316 & 0.249 & 0.527 & 0.777 \\
      hltcoe\cite{participants-hltcoe} &                    patapscoBM25RM3td* &    \xmark &   Sparse &      Org &          E &     \xmark &   0.315 & 0.318 & 0.243 & 0.509 & 0.792 \\
      hltcoe\cite{participants-hltcoe} &                patapscoBM25noRM3desc* &    \xmark &   Sparse &       MT &          E &     \xmark &   0.307 & 0.289 & 0.244 & 0.517 & 0.725 \\
      hltcoe\cite{participants-hltcoe} &                  patapscoBM25RM3desc* &    \xmark &   Sparse &      Org &          E &     \xmark &   0.300 & 0.297 & 0.241 & 0.473 & 0.764 \\
  naverloo\cite{participants-naverloo} &                    SpladeMiraclMonoqt &    \xmark &    Dense &      Org &          E &     \xmark &   0.291 & 0.297 & 0.205 & 0.408 & 0.665 \\
      hltcoe\cite{participants-hltcoe} &               patapscoBM25noRM3title* &    \xmark &   Sparse &      Org &          E &     \xmark &   0.280 & 0.267 & 0.186 & 0.435 & 0.638 \\
      hltcoe\cite{participants-hltcoe} &                 patapscoBM25RM3title* &    \xmark &   Sparse &      Org &          E &     \xmark &   0.279 & 0.261 & 0.203 & 0.460 & 0.714 \\
      hltcoe\cite{participants-hltcoe} &                patapscoBM25noRM3desc* &    \xmark &   Sparse &      Org &          E &     \xmark &   0.276 & 0.272 & 0.201 & 0.449 & 0.667 \\
\bottomrule
\end{tabular}
\begin{flushleft}
\footnotesize{Team \texttt{h2oloo} is renamed as \texttt{naverloo} to reflect the participating parties.}
\end{flushleft}

\end{table*}

%% file: _table_mlir_full_results.tex
\begin{table*}[]
\setlength\tabcolsep{0.5em}

\caption{MLIR Results.
The run used as the first stage retrieval for the reranking task is marked in bold. 
* indicates manual runs.
Column ``JFD'' indicates whether the run is \underline{j}udged at \underline{f}ull \underline{d}epth, which is 50. 
}\label{tab:mlir-full-results}
    \centering
\begin{tabular}{ll|ccccc|cccccc}
\toprule
                                Team &                    Run Name &      ReR. &    Model &       DL &         QL &        JFD &    nDCG & $\alpha$-nDCG &   RBP &    AP & R@100 &  R@1k \\
\midrule
naverloo\cite{participants-naverloo} &                       rgpt4 &    \xmark &   Hybrid &   MT+Org &          E &     \cmark &   0.523 &         0.789 & 0.694 & 0.434 & 0.564 & 0.881 \\
naverloo\cite{participants-naverloo} &                      frgpt4 &    \xmark &   Hybrid &   MT+Org &          E &     \cmark &   0.513 &         0.771 & 0.683 & 0.427 & 0.540 & 0.881 \\
naverloo\cite{participants-naverloo} &         RERANKEverythingRun &    \xmark &   Hybrid &   MT+Org &          E &     \cmark &   0.490 &         0.752 & 0.646 & 0.418 & 0.565 & 0.881 \\
naverloo\cite{participants-naverloo} &          RERANKBM25sSplades &    \xmark &   Hybrid &   MT+Org &          E &     \cmark &   0.485 &         0.753 & 0.645 & 0.416 & 0.566 & 0.877 \\
naverloo\cite{participants-naverloo} &                 RERANKBM25s &    \xmark &   Hybrid &   MT+Org &          E &     \cmark &   0.478 &         0.738 & 0.635 & 0.404 & 0.559 & 0.831 \\
naverloo\cite{participants-naverloo} &               EverythingRun &    \xmark &   Hybrid &   MT+Org &          E &     \cmark &   0.401 &         0.646 & 0.560 & 0.335 & 0.492 & 0.855 \\
        CIIR\cite{participants-CIIR} &           TransFuisonTrec23 &    \xmark &   Hybrid &       MT &          E &     \cmark &   0.389 &         0.578 & 0.557 & 0.345 & 0.497 & 0.852 \\
        CIIR\cite{participants-CIIR} &          HybridFuisonTrec23 &    \xmark &   Hybrid &       MT &          E &     \cmark &   0.388 &         0.584 & 0.560 & 0.336 & 0.503 & 0.840 \\
naverloo\cite{participants-naverloo} &                BM25sSplades &    \xmark &   Hybrid &   MT+Org &          E &     \cmark &   0.387 &         0.582 & 0.526 & 0.315 & 0.476 & 0.842 \\
    hltcoe\cite{participants-hltcoe} &                   colbertX* &    \xmark &    Dense &      Org &          E &     \cmark &   0.362 &         0.588 & 0.521 & 0.295 & 0.466 & 0.771 \\
    hltcoe\cite{participants-hltcoe} &       plaid\_v1\_mtt\_1bit* &    \xmark &    Dense &      Org &          E &     \cmark &   0.359 &         0.568 & 0.514 & 0.297 & 0.462 & 0.780 \\
    hltcoe\cite{participants-hltcoe} &          plaid\_v2\_eng\_1* &    \xmark &    Dense &       MT &          E &     \cmark &   0.355 &         0.530 & 0.511 & 0.319 & 0.491 & 0.804 \\
        CIIR\cite{participants-CIIR} &                      SPLADE &    \xmark & L-Sparse &       MT &          E &     \xmark &   0.355 &         0.569 & 0.503 & 0.301 & 0.463 & 0.804 \\
        CIIR\cite{participants-CIIR} &          NativeFuisonTrec23 &    \xmark &   Hybrid &      Org &          E &     \cmark &   0.350 &         0.560 & 0.516 & 0.280 & 0.438 & 0.772 \\
    hltcoe\cite{participants-hltcoe} & plaid\_v1\_mtt\_1bit\_date* &    \xmark &    Dense &      Org &          E &     \cmark &   0.335 &         0.532 & 0.483 & 0.266 & 0.416 & 0.720 \\
    hltcoe\cite{participants-hltcoe} &     \bf{patapscoBM25RM3td*} &    \xmark &   Sparse &       MT &          E &     \cmark &   0.306 &         0.519 & 0.470 & 0.272 & 0.430 & 0.764 \\
naverloo\cite{participants-naverloo} &                       BM25s &    \xmark &   Sparse &   MT+Org &          E &     \cmark &   0.303 &         0.507 & 0.423 & 0.234 & 0.396 & 0.782 \\
    hltcoe\cite{participants-hltcoe} &                  PSQraw-td* &    \xmark &   Sparse &      Org &          E &     \cmark &   0.295 &         0.437 & 0.427 & 0.233 & 0.403 & 0.693 \\
    hltcoe\cite{participants-hltcoe} &        patapscoBM25noRM3td* &    \xmark &   Sparse &       MT &          E &     \xmark &   0.290 &         0.499 & 0.431 & 0.236 & 0.397 & 0.721 \\
    hltcoe\cite{participants-hltcoe} &       patapscoBM25RM3title* &    \xmark &   Sparse &       MT &          E &     \xmark &   0.288 &         0.470 & 0.436 & 0.254 & 0.420 & 0.750 \\
    hltcoe\cite{participants-hltcoe} &        patapscoBM25RM3desc* &    \xmark &   Sparse &       MT &          E &     \xmark &   0.282 &         0.471 & 0.421 & 0.250 & 0.402 & 0.712 \\
    hltcoe\cite{participants-hltcoe} &     patapscoBM25noRM3title* &    \xmark &   Sparse &       MT &          E &     \xmark &   0.261 &         0.456 & 0.393 & 0.211 & 0.362 & 0.682 \\
    hltcoe\cite{participants-hltcoe} &                   PSQraw-t* &    \xmark &   Sparse &      Org &          E &     \xmark &   0.256 &         0.392 & 0.368 & 0.213 & 0.376 & 0.667 \\
    hltcoe\cite{participants-hltcoe} &      patapscoBM25noRM3desc* &    \xmark &   Sparse &       MT &          E &     \xmark &   0.251 &         0.434 & 0.374 & 0.208 & 0.362 & 0.662 \\

\bottomrule
\end{tabular}
\begin{flushleft}
\footnotesize{Team \texttt{h2oloo} is renamed as \texttt{naverloo} to reflect the participating parties.}
\end{flushleft}

\end{table*}

%% file: _table_tech_full_results.tex
\begin{table*}[]
\caption{Technical Document Task Results.
Monolingual runs, which use human translations of the queries, are shown in green.
The run used as the first stage retrieval for the reranking task is marked in bold. 
* indicates manual runs. All runs are judged with a depth of 20. 
}\label{tab:tech-full-results}
    \centering
\begin{tabular}{ll|cccc|ccccc}
\toprule
                                         Team &                          Run Name &      ReR. &    Model &       DL &         QL &    nDCG &   RBP &    AP & R@100 &  R@1k \\
\midrule
         naverloo\cite{participants-naverloo} &                 EverythingRun\_RR &    \xmark &   Hybrid &   MT+Org &          E &   0.496 & 0.429 & 0.391 & 0.764 & 0.944 \\
         naverloo\cite{participants-naverloo} &                EverythingRun\_fRR &    \xmark &   Hybrid &   MT+Org &          E &   0.493 & 0.428 & 0.387 & 0.781 & 0.944 \\
         naverloo\cite{participants-naverloo} &                  BM25sSplades\_RR &    \xmark &   Hybrid &   MT+Org &          E &   0.492 & 0.426 & 0.384 & 0.755 & 0.925 \\
         naverloo\cite{participants-naverloo} &                 BM25sSplades\_fRR &    \xmark &   Hybrid &   MT+Org &          E &   0.487 & 0.424 & 0.380 & 0.781 & 0.925 \\
         naverloo\cite{participants-naverloo} &                         BM25s\_RR &    \xmark &   Hybrid &   MT+Org &          E &   0.484 & 0.427 & 0.378 & 0.734 & 0.882 \\
         naverloo\cite{participants-naverloo} &                        BM25s\_fRR &    \xmark &   Hybrid &   MT+Org &          E &   0.479 & 0.416 & 0.369 & 0.760 & 0.882 \\
         naverloo\cite{participants-naverloo} &                     EverythingRun &    \xmark &   Hybrid &   MT+Org &          E &   0.430 & 0.388 & 0.317 & 0.668 & 0.944 \\
             hltcoe\cite{participants-hltcoe} &    \ml{plaid\_monozh\_mt5ht\_td*} &    \xmark &    Dense &      Org &          C &   0.410 & 0.378 & 0.308 & 0.639 & 0.812 \\
         naverloo\cite{participants-naverloo} &                      BM25sSplades &    \xmark &   Hybrid &   MT+Org &          E &   0.400 & 0.363 & 0.293 & 0.658 & 0.925 \\
             hltcoe\cite{participants-hltcoe} &                rerank\_mt5gt\_td* &    \cmark &   Hybrid &      Org &          E &   0.394 & 0.370 & 0.287 & 0.575 & 0.656 \\
             hltcoe\cite{participants-hltcoe} &             plaid\_tt\_mt5gt\_td* &    \xmark &    Dense &      Org &          E &   0.383 & 0.346 & 0.278 & 0.593 & 0.755 \\
             hltcoe\cite{participants-hltcoe} &           \ml{rerank\_mt5ht\_td*} &    \cmark &   Hybrid &      Org &          C &   0.378 & 0.366 & 0.288 & 0.612 & 0.722 \\
ISI\_SEARCHER\cite{participants-ISI_SEARCHER} &            run\_tech\_rr\_combine &    \xmark &   Hybrid &      Org &          E &   0.371 & 0.342 & 0.256 & 0.530 & 0.773 \\
ISI\_SEARCHER\cite{participants-ISI_SEARCHER} &                   run\_tech\_base &    \xmark &   Hybrid &      Org &          E &   0.362 & 0.320 & 0.241 & 0.443 & 0.773 \\
             hltcoe\cite{participants-hltcoe} &             plaid\_distilled\_td* &    \xmark &    Dense &      Org &          E &   0.360 & 0.334 & 0.253 & 0.555 & 0.824 \\
             hltcoe\cite{participants-hltcoe} &             \ml{plaid\_mono\_td*} &    \xmark &    Dense &      Org &          C &   0.359 & 0.321 & 0.231 & 0.509 & 0.758 \\
         naverloo\cite{participants-naverloo} &                            bm25qt &    \xmark &   Sparse &      Org &          E &   0.356 & 0.314 & 0.277 & 0.599 & 0.865 \\
ISI\_SEARCHER\cite{participants-ISI_SEARCHER} &        run\_tech\_rr\_combine\_td &    \xmark &   Hybrid &      Org &          E &   0.355 & 0.329 & 0.247 & 0.545 & 0.773 \\
             hltcoe\cite{participants-hltcoe} &  \ml{patapsco\_bm25\_ht\_t\_rm3*} &    \xmark &   Sparse &      Org &          C &   0.350 & 0.328 & 0.279 & 0.577 & 0.812 \\
         naverloo\cite{participants-naverloo} &                             BM25s &    \xmark &   Sparse &   MT+Org &          E &   0.345 & 0.306 & 0.258 & 0.602 & 0.882 \\
                 CIIR\cite{participants-CIIR} &                 TransFuisonTrec23 &    \xmark &   Hybrid &       MT &          E &   0.341 & 0.309 & 0.235 & 0.543 & 0.839 \\
         naverloo\cite{participants-naverloo} &                     mContrieverqt &    \xmark &    Dense &      Org &          E &   0.339 & 0.328 & 0.216 & 0.489 & 0.789 \\
             hltcoe\cite{participants-hltcoe} & \ml{patapsco\_bm25\_ht\_td\_rm3*} &    \xmark &   Sparse &      Org &          C &   0.331 & 0.320 & 0.266 & 0.585 & 0.813 \\
                 CIIR\cite{participants-CIIR} &                HybridFuisonTrec23 &    \xmark &   Hybrid &       MT &          E &   0.327 & 0.305 & 0.220 & 0.557 & 0.839 \\
         naverloo\cite{participants-naverloo} &                      SpladePPSDdt &    \xmark & L-Sparse &       MT &          E &   0.326 & 0.286 & 0.213 & 0.491 & 0.778 \\
             hltcoe\cite{participants-hltcoe} &       patapsco\_bm25\_qt\_t\_rm3* &    \xmark &   Sparse &      Org &          E &   0.322 & 0.320 & 0.246 & 0.556 & 0.793 \\
             hltcoe\cite{participants-hltcoe} &                     psq\_td\_f32* &    \xmark &   Sparse &      Org &          E &   0.314 & 0.296 & 0.206 & 0.477 & 0.768 \\
             hltcoe\cite{participants-hltcoe} &      patapsco\_bm25\_qt\_td\_rm3* &    \xmark &   Sparse &      Org &          E &   0.314 & 0.306 & 0.234 & 0.547 & 0.814 \\
         naverloo\cite{participants-naverloo} &                SpladeMiraclMonoqt &    \xmark & L-Sparse &      Org &          E &   0.311 & 0.296 & 0.202 & 0.440 & 0.743 \\
             hltcoe\cite{participants-hltcoe} &                      psq\_t\_f32* &    \xmark &   Sparse &      Org &          E &   0.310 & 0.279 & 0.204 & 0.468 & 0.760 \\
         naverloo\cite{participants-naverloo} &                   RetroMAEReprodt &    \xmark &    Dense &       MT &          E &   0.295 & 0.246 & 0.188 & 0.462 & 0.791 \\
         naverloo\cite{participants-naverloo} &                  SpladeMiraclENdt &    \xmark & L-Sparse &       MT &          E &   0.292 & 0.276 & 0.195 & 0.480 & 0.796 \\
             hltcoe\cite{participants-hltcoe} &       patapsco\_bm25\_qt\_d\_rm3* &    \xmark &   Sparse &      Org &          E &   0.290 & 0.287 & 0.218 & 0.501 & 0.755 \\
ISI\_SEARCHER\cite{participants-ISI_SEARCHER} &                     run\_tech\_rr &    \xmark &   Hybrid &      Org &          E &   0.290 & 0.272 & 0.198 & 0.496 & 0.773 \\
             hltcoe\cite{participants-hltcoe} &                   colbert\_x\_td* &    \xmark &    Dense &      Org &          E &   0.287 & 0.254 & 0.175 & 0.443 & 0.693 \\
             hltcoe\cite{participants-hltcoe} &                    plaid\_tt\_td* &    \xmark &    Dense &      Org &          E &   0.279 & 0.257 & 0.188 & 0.454 & 0.717 \\
             hltcoe\cite{participants-hltcoe} &  \ml{patapsco\_bm25\_ht\_d\_rm3*} &    \xmark &   Sparse &      Org &          C &   0.276 & 0.274 & 0.213 & 0.492 & 0.722 \\
             hltcoe\cite{participants-hltcoe} &               plaid\_V2model\_td* &    \xmark &    Dense &       MT &          E &   0.273 & 0.228 & 0.184 & 0.490 & 0.749 \\
             hltcoe\cite{participants-hltcoe} &                          blade-d* &    \xmark & L-Sparse &      Org &          E &   0.260 & 0.237 & 0.163 & 0.399 & 0.716 \\
             hltcoe\cite{participants-hltcoe} &                         blade-td* &    \xmark & L-Sparse &      Org &          E &   0.246 & 0.230 & 0.155 & 0.409 & 0.717 \\
                   AI2\cite{participants-AI2} &          \ml{Specter2Description} &    \xmark &    Dense &       MT &          E &   0.240 & 0.228 & 0.171 & 0.481 & 0.815 \\
             hltcoe\cite{participants-hltcoe} &                          blade-t* &    \xmark & L-Sparse &      Org &          E &   0.240 & 0.240 & 0.158 & 0.394 & 0.724 \\
                   AI2\cite{participants-AI2} &                \ml{Specter2Title} &    \xmark &    Dense &       MT &          E &   0.237 & 0.207 & 0.159 & 0.459 & 0.799 \\
         naverloo\cite{participants-naverloo} &                   SpladeNeuclirqt &    \xmark & L-Sparse &      Org &          E &   0.234 & 0.233 & 0.143 & 0.410 & 0.712 \\
                 CIIR\cite{participants-CIIR} &                NativeFuisonTrec23 &    \xmark &   Hybrid &      Org &          E &   0.231 & 0.218 & 0.148 & 0.405 & 0.745 \\
         naverloo\cite{participants-naverloo} &                            bm25dt &    \xmark &   Sparse &       MT &          E &   0.229 & 0.224 & 0.160 & 0.420 & 0.706 \\
                   AI2\cite{participants-AI2} &            \ml{Specter2Narrative} &    \xmark &    Dense &       MT &          E &   0.226 & 0.211 & 0.164 & 0.453 & 0.775 \\
             hltcoe\cite{participants-hltcoe} &           patapsco\_bm25\_t\_rm3* &    \xmark &   Sparse &       MT &          E &   0.222 & 0.199 & 0.142 & 0.371 & 0.660 \\
             hltcoe\cite{participants-hltcoe} &          patapsco\_bm25\_td\_rm3* &    \xmark &   Sparse &       MT &          E &   0.210 & 0.194 & 0.139 & 0.350 & 0.656 \\
             hltcoe\cite{participants-hltcoe} &           patapsco\_bm25\_d\_rm3* &    \xmark &   Sparse &       MT &          E &   0.189 & 0.173 & 0.123 & 0.354 & 0.616 \\
             hltcoe\cite{participants-hltcoe} &                plaid\_jhpolo\_td* &    \xmark &    Dense &      Org &          E &   0.072 & 0.074 & 0.036 & 0.099 & 0.212 \\
\bottomrule
\end{tabular}
\begin{flushleft}
\footnotesize{Team \texttt{h2oloo} is renamed as \texttt{naverloo} to reflect the participating parties.}
\end{flushleft}

\end{table*}